\documentclass[%
 reprint,
superscriptaddress,
 amsmath,amssymb,
 aps
]{revtex4-2}

\usepackage[colorlinks=true,linkcolor=blue,citecolor=blue,urlcolor=blue]{hyperref}
\usepackage{graphicx}
\usepackage{dcolumn}
\usepackage{amsmath,amssymb,physics}
\usepackage{cleveref}
\usepackage{multirow}
\usepackage{array}
\usepackage{mathtools}
\usepackage{orcidlink}

\graphicspath{{./figures/}}

\usepackage{float}
\usepackage{xcolor}
\usepackage{booktabs}
\usepackage{subcaption}
\usepackage{lineno}

\usepackage{etoolbox}
\makeatletter
\pretocmd\@makecaption{\nolinenumbers}{}{}
\apptocmd\@makecaption{\linenumbers}{}{}
\makeatother

\usepackage{comment}
\usepackage{bbold, mathrsfs, slashed}

\usepackage{titlesec}

\begin{document}

\preprint{APS/123-QED}

\title{Analyzing the dense matter equation of states in the light of the compact object HESS J1731-347 } 

\author{Skund Tewari \orcidlink{[0009-0004-2420-7935}}
\email{skund20@iiserb.ac.in}
\affiliation{Indian Institute of Science Education and Research Bhopal, Bhopal 462066, India}

\author{Sagnik Chatterjee \orcidlink{0000-0001-6367-7017} }
\email{sagnik18@iiserb.ac.in}
\affiliation{Indian Institute of Science Education and Research Bhopal, Bhopal 462066, India}

\author{Deepak Kumar \orcidlink{0000-0001-9292-3598}}
\email{deepak@iiserb.ac.in}
\affiliation{Indian Institute of Science Education and Research Bhopal, Bhopal 462066, India}
\affiliation{Institute of Physics, Sachivalaya Marg, Bhubaneswar 751005, India}

\author{Ritam Mallick \orcidlink{0000-0003-2943-6388}}
\email{mallick@iiserb.ac.in}
\affiliation{Indian Institute of Science Education and Research Bhopal, Bhopal 462066, India}

\date{\today}

\begin{abstract}
The recent mass ($0.77 \pm ^{0.20}_{0.17}M_{\odot}$) and radius ($10.4\pm^{0.86}_{0.78} \text{km}$) measurement of HESS J1731-347 made it one of the most fascinating object if it is indeed a neutron star. In this work, we examine the current status of the dense matter equation of states in the context of this compact object being a neutron star. We use three sets of equation of states corresponding to the three classes - neutron stars, strange stars, and hybrid stars and perform Bayesian model selection on them. Our results show that for hadronic models, the EoS is preferred to be stiff at the intermediate densities. This makes the Brueckner-Hartree-Fock approximation and models based on effective interactions deviate from current astrophysical observations on the inclusion of HESS J1731-347. Furthermore, for the strange star family, the equation of states composed of three flavor quarks prefers relatively smaller bag parameters. Analyzing the hybrid family of equation of states consisting of a first-order phase transition revealed preferences for early first-order phase transition. Comparing all the preferred equations of state among each family, it was found that the current astrophysical constraints prefer the hybrid equation of states the most.
\end{abstract}
\maketitle

\section{Introduction} \label{sec:intro}
\vspace{-5pt}

Quantum chromodynamics (QCD) predicts the existence of quarks and gluons in a deconfined state at higher densities \citep{Fraga_2014, Fadafan_2020} hinting towards a possible phase transition (PT) at some intermediate densities  \citep{Iso_2017}. The dense core of neutron stars (NSs) lies at the lower end of the intermediate densities (typical central densities lie in the range of 2-8 times that of the nuclear saturation density ($n_{s} = 0.16 \:\text{fm}^{-3}$), making them one of the most fascinating compact objects to study \citep{glendenning2012compact}. Terrestrial-based laboratories are yet to reach such densities, further fuelling our interest in these extreme objects. 

The core of the NSs is still a mystery, with several models suggesting the presence of only hadronic matter \citep{Bednarek, Xia_2022, rather2022}. The possibility of PT at intermediate densities opens the possibility of hybrid stars \citep{Hybrid_Alford_2005,Hybrid_DRAGO2001229,Hybrid_Anil,Hybrid_Benhar,Hybrid_Li_2023,PT_LaskosPatkos}, having an inner core composed of quark matter followed by an outer core of hadronic matter \citep{Bhattacharyya:2006vy, Sinha, Mallick:2020bdc, Kuzur:2021gnf, Gorda_2023, Thakur:2024ijp}. There also exists another unique category of stars called strange stars (SSs) made up of strange quark matter (SQM) \citep{Weber_2012, Nandi:2020luz, Anil_quark}. Witten suggested that the SQM is the absolute ground state consisting of $u$,$d$, and $s$ quarks supporting the idea of SSs \citep{Witten}.

The only way to demystify the core of NSs is with astrophysical observations. Apart from the mass measurements of PSR J0348+0432, we also have the mass measurement of PSR J0740+6620 \citep{Fonseca:2021} which is $2.08\pm 0.7 M_{\odot}$. Simultaneous mass and radius measurements from NICER of PSR J0740+6620 \citep{Miller_2021, Riley_2021} and PSR J0030+0451 \citep{Miller_2019, Riley_2019} have helped in constraining the EoSs. The EoSs have been further constrained from the binary tidal deformability measurement ($\Tilde{\Lambda} < 720$) imposed by the GW170817 event of the binary neutron star merger reported by the LIGO collaboration \citep{Abbott_2017a, Abbott_2017b, Abbott_2018}. These observations have been successful in narrowing down the EoS band constructed from model agnostic approaches \citep{Annala_2020, Altiparmak_2022}. With more improved observations in the future, the EoS band is expected to get thinner. 

However, the recent observation of HESS J1731-347 \citep{Doroshenko2022} started raising a few eyebrows with a mass and radius measurement of $0.77 \pm ^{0.20}_{0.17} M_{\odot}$ and  $10.4\pm^{0.86}_{0.78} \text{km}$ respectively, making it one of the lightest known compact stars till date. {A recent work highlights that the analysis of the central compact object (CCO) in the supernova remnant (SNR) HESS J1731-347 is dependent on a number of necessary but insufficient conditions \citep{Alford_2023_CCO_UTCA}. In particular, the authors found that the assumption of a uniform temperature carbon atmosphere (UTCA) adopted by Ref. \citep{Doroshenko2022} is not at all consistent with the longest \& highest quality XMM-Newton data, further stating that a UTCA model for the CCO is less likely.

Although these points portray that the mass-radius calculation of the CCO is controversial, another recent work by Ref \citep{HESS_Consistency} talks about the minimal consistency checks for the observation of the remnant in HESS J1731-347 within existing models of SSs. They concluded with the remark that the CCO in the supernova remnant HESS J1731-347 passes initial consistency checks and can be utilised for theoretical and observational work. Furthermore, the recent work of Ref. \citep{zhang2024_HESS_lightest} explores the possibility of HESS J1731-347 indeed being the lightest star observed. Their analysis found evidence that the star could indeed be a NS of mass 0.9 M$_{\odot}$ formed from the collapse of a fast-rotating iron core. It is, therefore, necessary to at least check its effect on the present EoS bounds set by NS observations.
}

Several significant works followed this observation, each of them trying to explain the properties of this star \citep{Sagun_2023, Rather_2023, Kubis,HybridHESS, veselsky2024,LaskosPatkos_HESS,Brodie_HybridHESS,Oikonomou_QM_HESS}. 
However, one of the best ways to examine any observation is to do a statistical analysis with the data (basically different models) one has. Bayesian analysis is an important tool that is being used significantly to constrain nuclear models \citep{Tuhin1, Tuhin2, Tuhin3}.
Recently, Ref \cite{BhaskarHESS} showed, using a Bayesian framework, the impact of the compact object in the context of nuclear matter.

Bayesian inference techniques have been primarily used for constraining GW observation parameters \citep{Abbott_2020b, Ghosh_2021}, with Ref \cite{Biswas_2022} developing a model selection technique using various astrophysical observations to compare nuclear matter EoSs. {Similarly, another paper presents the application of Bayesian model selection, ranking a few EoSs using only the GW data of GW170817 \citep{Pacilio_OnlyGW_Bayesian}. By using only GW data, one can only constrain the equation of state at densities corresponding to the central density of canonical neutron star of mass 1.4M$_{\odot}$.} In this paper, we perform Bayesian model selection to explore the implications of the compact object HESS J1731-347, along with other astrophysical observations in the light of a neutron star, a strange star, or a hybrid star. {Since SSs inherently favor low mass and radius, we cannot rule out the possibility of such stars. Additionally, by using several astrophysical observations, we can constrain the EoS by surveying a wide range of densities. Furthermore,}  we consider several nuclear matter EoS models, models consisting of three flavored quarks, and hybrid models based on first-order phase transition (FOPT). Lastly, we compare the three sets of EoSs to analyze which family of EoSs explains the current astrophysical observations the best.

The paper is arranged in the following way. Section \ref{sec:formalism} discusses the formalism adopted in this work to describe the EoS models and also the model selection technique and how it is implemented in this work. The results are described in section \ref{sec:Results}, and finally, in section \ref{sec:Summ}, we conclude with a summary and discussion of our results.

\section{Formalism} \label{sec:formalism}
\vspace{-5pt}

{In order to use the EoSs for astrophysical analysis, we first checked whether the EoSs satisfied constraints from chiral effective field theory. To do so, we adopted a construction similar to Refs. \citep{Altiparmak_2022,Pratik_CETcostruction}, where, for densities in the range $0<n \leq0.5n_s$, where $n_s$=0.16 fm$^{-3}$, the Baym-Pethick-Sutherland (BPS) \citep{QMC_BPSA} EoS is considered. In the density range of $0.5n_s<n \leq1.1n_s$, a series of monotropes of the form $p=K\rho^{\gamma}$ are constructed, where the value of `K' is fixed by matching to the BPS EoS and $\gamma$ is varied in the range [1.77,3.23]. These different monotropes in the density range $0.5n_s<n\leq1.1n_s$ form our CET band. We then check whether all the EoSs that we use for our analysis lie within this region. Our analysis did not use any EoS that did not comply with this condition. In Appendix \ref{appx:noCET}, we discuss the changes in our study when the CET constraint is not taken into consideration.}

\subsection{Hadronic EoSs} \label{subsec:had}
For the purpose of hadronic EoSs, we use the publicly available nuclear matter EoSs in the {\ttfamily CompOSE} repository \citep{compose_website}. We have considered 7 EoS models, namely the Density-dependent Relativistic Mean Field (RMF) model, the Effective Interaction model, the Thomas Fermi approach, the RMF model, the Brussels-Montreal energy density functional, non-linear RMF model, and the Brueckner-Hartree-Fock approximation. Along with these, we have also included APR(APR) (an EoS based on variational techniques), CMGO(GDFM-I) (an EoS based on density-dependent covariant density functional), and PT(GRDF2-DD2) (an EoS based on generalized relativistic density functional). More details about the EoSs used can be found in \cref{tab:had}.

\begin{table*}
    \centering
    \caption{\raggedright \small{List of all the 38 hadronic EoS with the corresponding radius of a $0.77M_{\odot}$, $1.4M_{\odot}$, and $2M_{\odot}$ star, respectively; along with the Tidal Deformability measurement of the $1.4M_{\odot}$ star and the maximum mass given by each EoS. The last three equations are based on variational techniques, density-dependent covariant density functional, and generalized relativistic density functional, respectively. The information of the rejected hadronic EoSs is presented in bold text.}}
    \resizebox{\textwidth}{!}{
    \begin{tabular}{>{\raggedright}p{9.29cm}c@{\hspace{0.85cm}} c@{\hspace{.85cm}} c@{\hspace{0.85cm}} c@{\hspace{0.85cm}} c c}
    \toprule\toprule
       {EoS}  & {$R_{0.77}$(km)} & {$R_{1.4}$(km)} & {$R_{2.0}$(km)} & {$\Lambda_{1.4}$} & {$M_{max}(\text{M}_{\odot})$} \\
        \midrule 
        \multicolumn{6}{l}{\textit{{Density Dependent RMF model}}} \vspace{0.5pt} \\ 
        {GPPVA(TW)NSunifiedInnerCrust-core} \citep{Grill,TW1999} & \textbf{12.75} & \textbf{12.33} & \textbf{11.42} & \textbf{401}  & \textbf{2.07}\\
        {SPG(M2)unifiedNSEoS} \citep{PCP,CMGO} & 12.45 & 12.63 & 12.58 & 518  & 2.42\\
        {SPG(M4)unifiedNSEoS} \citep{PCP,SPG} & 12.18 & 12.31 & 12.22 & 433  & 2.35\\
        {SPG(M5)unifiedNSEoS} \citep{PCP} & 13.15 & 13.42 & 13.65 & 772  &2.71 \\
        {SPG(M3)unifiedNSEoS} \citep{PCP} & 12.44 & 12.65 & 12.95 & 523   & 2.69\\
        {GPPVA(DD2)NSunifiedInnerCrust-core} \citep{GRDF2-DD2} & 13.06 & 13.19 & 13.14 & 683  &2.42 \\
        {SPG(M1)unifiedNSEoS} \citep{PCP} & 12.78 & 12.8 & 12.87 & 534   & 2.54 \vspace{0.5pt} \\
        \multicolumn{6}{l}{\textit{{Effective Interactions}}} \vspace{0.5pt} \\
        {RG(SkMp)} \citep{RG,SkMp} & \textbf{12.54} & \textbf{12.5} & \textbf{11.5} & \textbf{467}  & \textbf{2.11}\\
        {RG(SkI4)} \citep{RG,SkI4_SKI3_SKI2_SKI6_SKI5} & 12.18 & 12.38 & 11.74 & 458  & 2.18\\
        {RG(SKb)} \citep{RG,SKb_SKa} & 11.82 & 12.21 & 11.69 & 404  & 2.2\\
        {RG(SLY2)} \citep{RG,SLY2_SLY9} & \textbf{11.93} & \textbf{11.79} & \textbf{10.7} & \textbf{307}  & \textbf{2.06}\\
        {RG(SLY230a)} \citep{RG,SLY230a} & \textbf{11.90} & \textbf{11.83} & \textbf{11.05} & \textbf{324} & \textbf{2.11}\\
        {RG(SKa)} \citep{RG,SKb_SKa} & \textbf{13.02} & \textbf{12.92} & \textbf{12.16} & \textbf{558}   & \textbf{2.22}\\
        {VGBCMR(D1MStar)} \citep{VGBCMR} & \textbf{11.67} & \textbf{11.71} & \textbf{10.47} & \textbf{314}  & \textbf{2.00}\\
        {RG(SLY9)} \citep{RG,SLY2_SLY9} & \textbf{12.53} & \textbf{12.47} & \textbf{11.70} & \textbf{444} & \textbf{2.16}\\
        {RG(SkI6)} \citep{RG,SkI4_SKI3_SKI2_SKI6_SKI5} & 12.33 & 12.49 & 11.88 & 481 & 2.2 \\
        {RG(SLY4)} \citep{RG,SLY4} & \textbf{11.84} & \textbf{11.7} & \textbf{10.62} & \textbf{295} & \textbf{2.06} \vspace{0.5pt} \\
        \multicolumn{6}{l}{\textit{{Thomas Fermi approach}}} \vspace{0.5pt} \\
        {XMLSLZ(DD-LZ1)} \citep{Xia_2022,Wei_2020_DDLZ1} & 12.52 & 13.15 & 13.34 & 732 & 2.56\\
        {XMLSLZ(DDME2)} \citep{Xia_2022,DDME2} & 12.74 & 13.2 & 13.22 & 712  & 2.48\\
        {XMLSLZ(DDME-X)} \citep{Xia_2022,DDME-X} & 12.81 & 13.37 & 13.49 & 792 & 2.56\\
        {XMLSLZ(TW99)} \citep{Xia_2022,TW1999} & \textbf{12.35} & \textbf{12.27} & \textbf{11.35} & \textbf{405} & \textbf{2.08} \vspace{0.5pt}\\
        \multicolumn{6}{l}{\textit{{RMF approximation}}} \vspace{0.5pt} \\
        {PCGS(PCSB1)} \citep{Pradhan,PCSB_HTZCS} & \textbf{12.98} & \textbf{13.25} & \textbf{12.67} & \textbf{624} & \textbf{2.19}\\
        {PCGS(PCSB0)} \citep{Pradhan,PCSB_HEMPEL} & 13.04 & 13.3 & 13.28 & 713  & 2.53\\
        {ABHT(QMC-RMF2)} \citep{ABHT} & \textbf{12.00} & \textbf{12.03} & \textbf{11.02} & \textbf{354} & \textbf{2.04} \\
        {ABHT(QMC-RMF3)} \citep{ABHT,QMC_BPSA} & \textbf{12.33} & \textbf{12.26} & \textbf{11.61} & \textbf{386} & \textbf{2.15}\\
        {ABHT(QMC-RMF4)} \citep{ABHT,Grill} & 12.00 & 12.35 & 12.04 & 420 & 2.21 \\
        {PCP(BSK26)} \citep{PCP,BSK_AC} & 11.7 & 11.77 & 11.18 & 323 & 2.17 \vspace{0.5pt}\\
        \multicolumn{6}{l}{\textit{{Brussels-Montreal energy density functionals}}} \vspace{0.5pt} \\
        {PCP(BSK25)} \citep{PCP,BSK_PC} & 11.97 & 12.37 & 12.10 & 476 & 2.22\\
        {PCP(BSK24)} \citep{PCP,BSK_GCP} & 12.26 & 12.5 & 12.27 & 514   &2.28 \\
        {PCP(BSK22)} \citep{PCP,BSK_PCS} & 12.97 & 13.04 & 12.58 & 624 & 2.26 \vspace{0.5pt}\\
        \multicolumn{6}{l}{\textit{{Nonlinear RMF models}}} \vspace{0.5pt} \\
        {GPPVA(NL3wrL55)NSunifiedInnerCrust-core} \citep{NL3wrL55} & {13.32} & {13.76} & {14.06} & {939} & {2.75} \\
        {GPPVA(FSU2H)NSunifiedInnerCrust-core} \citep{FSU2H} & 12.91 & 13.29 & 10.26 & 750   &2.37 \\
        {GPPVA(TM1e)NSunifiedInnerCrust-core} \citep{TM1e} & \textbf{13.02} & \textbf{13.16} & \textbf{10.59} & \textbf{661}   & \textbf{2.12} \vspace{0.5pt}\\
        \multicolumn{6}{l}{\textit{{Brueckner-Hartree-Fock approximations}}} \vspace{0.5pt} \\
        {BL(chiral)withUnifiedCrust} \citep{Bombaci} & \textbf{12.60} & \textbf{12.27} & \textbf{11.13} & \textbf{386}  & \textbf{2.08}\\
        {BL(chiral)WithCrust} \citep{Bombaci,DHA_chiralwithcrust} & \textbf{12.62} & \textbf{12.31} & \textbf{11.13} & \textbf{385}   & \textbf{2.08} \vspace{0.5pt} \\
        APR(APR) \citep{APR} & 11.31 & 11.33 & 10.85 & 248 & 2.19 \\
        CMGO(GDFM-I) \citep{CMGO} & 12.72 & 12.81 & 12.46 & 533 & 2.31\\
        PT(GRDF2-DD2)coldNS \citep{PT,GRDF2-DD2} & 12.84 & 13.17 & 13.07 & 686 & 2.42\\
    \bottomrule\bottomrule
    \end{tabular}%
    }
    \label{tab:had}
\end{table*}

\subsection{Quark Matter EoSs} \label{subsec:quark_eos}

The density in the core of NS can reach a few times the nuclear saturation density. At such high densities, the quarks may gain asymptotic degrees of freedom rather than nucleons/hadrons. In the present study, we consider quark matter with $u$, $d$, and $s$ quarks and electron as the only lepton. We adopt a three flavors modified MIT bag model with quark-vector meson interaction, which regulates the stiffness/softness of an EoS \cite{Lopes:2020btp, Pal:2023dlv, Podder:2023dey}. This model has three free parameters: (i) the bag constant $B$, which is still an inclusive parameter and defines the pressure on the walls of the bag to balance the degeneracy pressure of quarks. It plays an important role in determining the properties of quark stars. Its numerical value is not fixed. Here, in the present study, we consider it in the huge range $\in [139,\ 150]$ MeV range. (ii) The scaled coupling constants $x_{\rm v} = \left( \frac{g_{u\omega}}{g_{s\omega}} \right)$ and a vector coupling constant $g_{\rm v} = \left( \frac{g_{u\omega}}{m_\omega}\right)^2$. We also consider that the couplings between $u,\ d$ quarks and vector meson $\omega$ are equal $g_{u\omega} = {g_{d\omega}} = \sqrt{g_{\rm v}}m_{\omega}$. The parameter $m_{\omega} = 782.5\ {\rm MeV}$ is the mass of vector $\omega$ meson. In the previous studies, Ref. \cite{Pal:2023dlv}, the different values of $x_{\rm v}$ are studied. In the present study we consider  $x_{\rm v} \in [0,\ 1]$ while keeping $g_{\rm v} = 0.3$ fm$^2$. (iii) The dimensionless self-interaction, $b_4$ coupling of vector meson $\omega$. This parameter is also important in determining the EoS of quark stars. The negative value of this parameter gives a stiffer EoS and, hence, a larger mass quark star, while its positive value shows opposite results. In Refs. \cite{Lopes:2020btp}, we can see that the value of the parameter $b_4 = -0.4$ gives a larger mass quark star and $b_4 = 1.0$ gives smaller mass quark star. In the present study, we choose $b_4 = -0.4$. The different combinations of these parameters result in different quark matter EoS, which we use to determine the first-order phase transition.

\subsection{EoSs with First-Order PT} \label{subsec:PT_EoS}
Assuming the phase transition to be a FOPT, we construct the hybrid EoS from the hadronic and modified MIT bag-model quark matter EoS. The jump/transition from the hadronic phase (HP) to the quark phase (QP) happens at a particular pressure when the chemical potential of the quark phase becomes less than the chemical potential of the hadronic phase. Although this occurs at a specific pressure and chemical potential, and they remain smooth throughout, there is a discontinuity in the energy density (and density) corresponding to the latent heat required for the transition. {First-order phase transition can be of two types depending on the surface tension between the adjoining fluids: Gibbs and Maxwell. The surface tension of the quark matter is the decisive parameter that dictates the type of phase transition construction mechanism. Although this parameter is poorly known, its theoretical estimates fall within a wide range of (5-300) MeV/fm$^{3}$. In Refs. \cite{Heiselberg:1992dx, Iida:1998pi} its value was found to lie in the range (10-50) MeV/fm$^{3}$. However, in Ref. \cite{Voskresensky:2002hu}, authors estimated its value in the range (50-150) MeV/fm$^{3}$ and in Ref. \cite{Alford:2001zr}, an even higher value was estimated. Since we do not know the correct value of surface tension, both scenarios can be utilized to construct the phase transition.}

In the present study, we have considered Maxwell's construction mechanism, with $p_{\rm HP}(\mu_{\rm c}) = p_{\rm QP}(\mu_{\rm c})$ (where $\mu_{\rm c}$ is the critical baryonic chemical potential where the transition occurs), {while considering a large value of surface tension.  It should be noted that the FOPT could have also been modeled using Gibbs construction. However, this construction requires information on the chemical potential component-wise. Since {\ttfamily CompOSE} does not provide this information, we utilize the overall chemical potential of the state, provided by {\ttfamily CompOSE}, to perform Maxwell construction.}

\subsection{Bayesian Model Selection} \label{sec:Bays}

We adopt a Bayesian model selection approach to compare various models of EoS. Each unique EoS is considered a model, and we use Bayes' theorem, defined as:

\begin{equation}
    P(M|d, I)=\dfrac{P(d|M, I)P(M|I)}{P(d|I)}
\end{equation}

where $M$ refers to a model (EoS), $I$ refers to any background information we have, and d refers to the astrophysical data. $P(M|d,I)$ is the posterior probability of the model, $P(d|M,I)$ is the marginalized-likelihood (evidence) for the data, $P(M|I)$ is the prior probability and $P(d|I)$ is a constant term. \\

The evidence value, $P(d|M,I)$, can be obtained by marginalizing over the parameters of the model as :
\begin{align}
    P(d|M, I) &=\int P(d, \theta |M, I) d\theta   \nonumber \\
    & = \int P(d|\theta , M, I)P(\theta |M, I) d\theta \label{evd}
\end{align}

where $\theta$ refers to the parameters of the model, $P(d|\theta,M, I)$ is the likelihood function of the parameters, and $P(\theta | M,I)$ is a prior probability on the parameters given the information of the model. The evidence value for a model is independent of other models and remains constant irrespective of the number of models evaluated simultaneously.

In order to compare two different models ($M_1$ and $M_2$), we find out the odds ratio between them, which is defined as:

\begin{equation}
    \mathcal{O}_{M_1}^{M_2}=\dfrac{P(M_2|d,I)}{P(M_1|d,I)}
    =\dfrac{P(d|M_{2},I)}{P(d|M_{1},I)}\times \frac{P(M_2|I)}{P(M_1|I)}
\end{equation}

where the ratio of the likelihood for the two models is known as the Bayes factor. For uniformity, we can take the ratio of the priors to be equal to one so that $P(M_2|I) = P(M_1|I)$. By doing so, we avoid the preference for one model over the other. This choice can be modified depending on the background information.

Hence, the odds ratio is redefined as:

\begin{equation}
    \mathcal{O}_{M_1}^{M_2}=\dfrac{P(d|M_{2},I)}{P(d|M_{1},I)}
\end{equation}

If the value of the odds ratio is much greater than 1, then model $M_2$ is preferred over model $M_1$. If the ratio is much smaller than 1, then the inverse is true.

If we perform the analysis for multiple datasets, such that $d=\{d_k\}$, then:

\begin{equation}
    P(\{d_k\}|M,I)=\prod_k P(d_k|M,I)
\end{equation}

The odds ratio then finally takes the form: 

\begin{align}
    \mathcal{O}_{M_1}^{M_2}= \prod_k \frac{P(d_k|M_{2},I)}{P(d_k|M_{1},I)}
\end{align}

For our analysis, $d=\{d_{\rm GW},\ d_{\rm HESS},\ d_{\rm NICER}\}$ refers to the three sets of astrophysical observations we have used. The mass and tidal deformability  ($\Lambda$) measurements from GW170817 \citep{Abbott_2017a, Abbott_2017b, Abbott_2019} serves as $d_{\rm GW}$. For $d_{\rm NICER}$, the mass and radius measurements from PSR J0030+0451 \citep{Miller_2019}, {PSR J0437-4715}, and PSR J0740+6620 \citep{Miller_2021} serve as input. Similarly, the mass and radius measurements from HESS J1731-347 \citep{Doroshenko2022} form $d_{\rm HESS}$. As both $d_{\rm HESS}$ and $d_{\rm NICER}$ consist only of the mass and radius measurements, the evidence calculation is the same for them. 

First, let us calculate the evidence for the mass-radius measurements for the PSR J0030+0451, PSR J0740+6620, and HESS J1731-347 observations. Since we consider each EoS as a model, we replace `$M$' with `EoS' in eq \eqref{evd}. For every EoS, we solve the Tolman-Oppenheimer-Volkoff (TOV) equations \citep{TOV} to obtain the mass-radius curve, also known as the MR curve. Our EoS can be parametrized either by using the mass or by the radius values obtained after solving the TOV equations. In our scenario, we use the mass values, and hence, the evidence for the NICER observations is given by:

\begin{align}
P(d_{\rm NICER}|{\rm EoS}, I)&= \:\int_{m_{\rm min}}^{m_{\rm max}}P(d_{\rm NICER}|m, \nonumber \\
& R(m,EoS),EoS,I) \nonumber \\ 
& \times P(m|EoS,I)dm
\end{align}

where $P(m|EoS,I)$ is the prior distribution on our parameter and $P(d_{\rm NICER}|m,R,EoS,I)$ is the likelihood of the data.

Similarly, for the HESS observation, the evidence is given as :

\begin{align}
P(d_{\rm HESS}|EoS, I)&= \:\int_{m_{\rm min}}^{m_{\rm max}}P(d_{\rm HESS}|m, \nonumber \\
&R(m,EoS),EoS,I) \nonumber \\ 
& \times P(m|EoS,I)dm
\end{align}
with $P(d_{\rm HESS}|m,R,EoS,I)$ being the likelihood of the data.

Without loss of generality, we can choose a uniform prior on mass \citep{Biswas_2022, BhaskarHESS}. It is given by:

\begin{equation}
    \begin{split}
        P(m|EoS, I)& = \frac{1}{m_{\rm max}-m_{\rm min}}; \: \text{$m_{\rm min}\leq m \leq m_{\rm max}$}\\
    & = 0 \hspace{2.2cm} ; \text{everywhere else} \label{prior}
    \end{split}
\end{equation}

$m_{max}$ is the maximum mass of the EoS obtained after solving the TOV equation. We fix $m_{min}$ equal to $0.5M_{\odot}$. To construct the likelihoods $P(d_{\rm NICER}|m,R,EoS,I)$ and $P(d_{\rm HESS}|m,R,EoS,I)$, we use a Gaussian kernel density estimation (KDE) with the mass and radius samples from NICER and HESS. 

To calculate the evidence for the GW data, we parameterize the two masses of binaries ($m_1$,$m_2$) and their corresponding tidal deformabilities ($\lambda_1$,$\lambda_2$) as:
\begin{align}
    P(d_{\rm GW}|EoS,I)&=\int_{m_2}^{M_{\rm max}} dm_1\int_{M_{\rm min}}^{m_1} P(d_{\rm GW}|m_1,m_2, \nonumber \\
    &\lambda_1(EoS,m_1),\lambda_2(EoS,m_2),EoS,I) \nonumber \\
    & \times P(m_1,m_2|EoS,I)dm_2 \label{dGW}
\end{align}

To solve eq \eqref{dGW} we make use of the chirp mass \citep{Raithel_2018} given by:

\begin{equation}
    \mathcal{M}_{\rm chirp} = \dfrac{(m_1m_2)^{3/5}}{(m_1+m_2)^{1/5}} = 1.186 M_{\odot}
\end{equation}

Where $m_1$ and $m_2$ are the masses of the primary and secondary neutron stars having a mass ratio, $q = {m_2}/{m_1} \ge 0.73$ inferred from GW170817 observation \citep{Abbott_2019}. Doing so reduces the parameters needed to evaluate the integral for the evidence of GW170817. We also use the same prior distribution as eq \eqref{prior} for the GW observation. We construct the likelihood using a multivariate Gaussian KDE with the mass and tidal deformability samples from the observation.

All of the evidence integrals were performed using {\ttfamily PyMultiNest} \citep{Buchner_Pymultinest}, which is a {\ttfamily Python} package for implementing the { \ttfamily MultiNest} algorithm. It offers efficient evidence calculation for multi-modal data. Furthermore, all likelihood distributions were constructed using the multivariate KDE method of { \ttfamily Statsmodels} \citep{seabold2010statsmodels}.


\section{Results} \label{sec:Results}
\vspace{-5pt}

For our analysis, the observations we have used are $(i)$ GW170817, $(ii)$ Three X-Ray sources, namely PSR J0030+0451\citep{Miller_2019}, {PSR J0437-4715}\citep{Choudhury_2024_PSRJ0437}, PSR J0740+6620 \citep{Miller_2021}, and HESS J1731-347 \citep{Doroshenko2022}. {Appendix \ref{appx:noHess} outlines the results of our analysis when the data of HESS J1731-347 is not taken into consideration.}

For the three sets of EoSs we have considered in our analysis, we have {38} hadronic EoSs, 58 strange matter EoSs, and {544} hybrid EoSs. Since the hybrid EoSs were constructed using Maxwell construction, for each nuclear EoS considered, we agnostically generated a family of hybrid EoSs, resulting in a large number of hybrid EoSs. We adopt Jeffrey's \citep{jeffreys1998probability} scale for the log of the odds ratio values. Jeffrey's scale is defined for the Bayes factor, not the odds ratio. However, upon taking the ratio of the prior of each model to be unity, the odds ratio becomes equal to the Bayes factor. According to the scale, if $\log_{10}\mathcal{O}_{M_1}^{M_2}$ lies between (-0.5, 0), then although there is evidence for model $M_1$, it is not worth more than a bare mention. If $\log_{10}\mathcal{O}_{M_1}^{M_2}$ lies between (-1, -0.5), then there is `substantial' evidence for (against) model $M_1$ ($M_2)$. If $\log_{10}\mathcal{O}_{M_1}^{M_2}$ lies between (-2, -1), then there is `strong' evidence for (against) model $M_1$ ($M_2$). If $\log_{10}\mathcal{O}_{M_1}^{M_2}$ is smaller than -2, then there is `decisive' evidence for (against) model $M_1$ ($M_2$), and we can reject model $M_2$. Utilizing this, we present our analysis in the following subsections.

\subsection{Hadronic EoS}
\begin{figure}
    \centering 
    \includegraphics[scale=0.43]{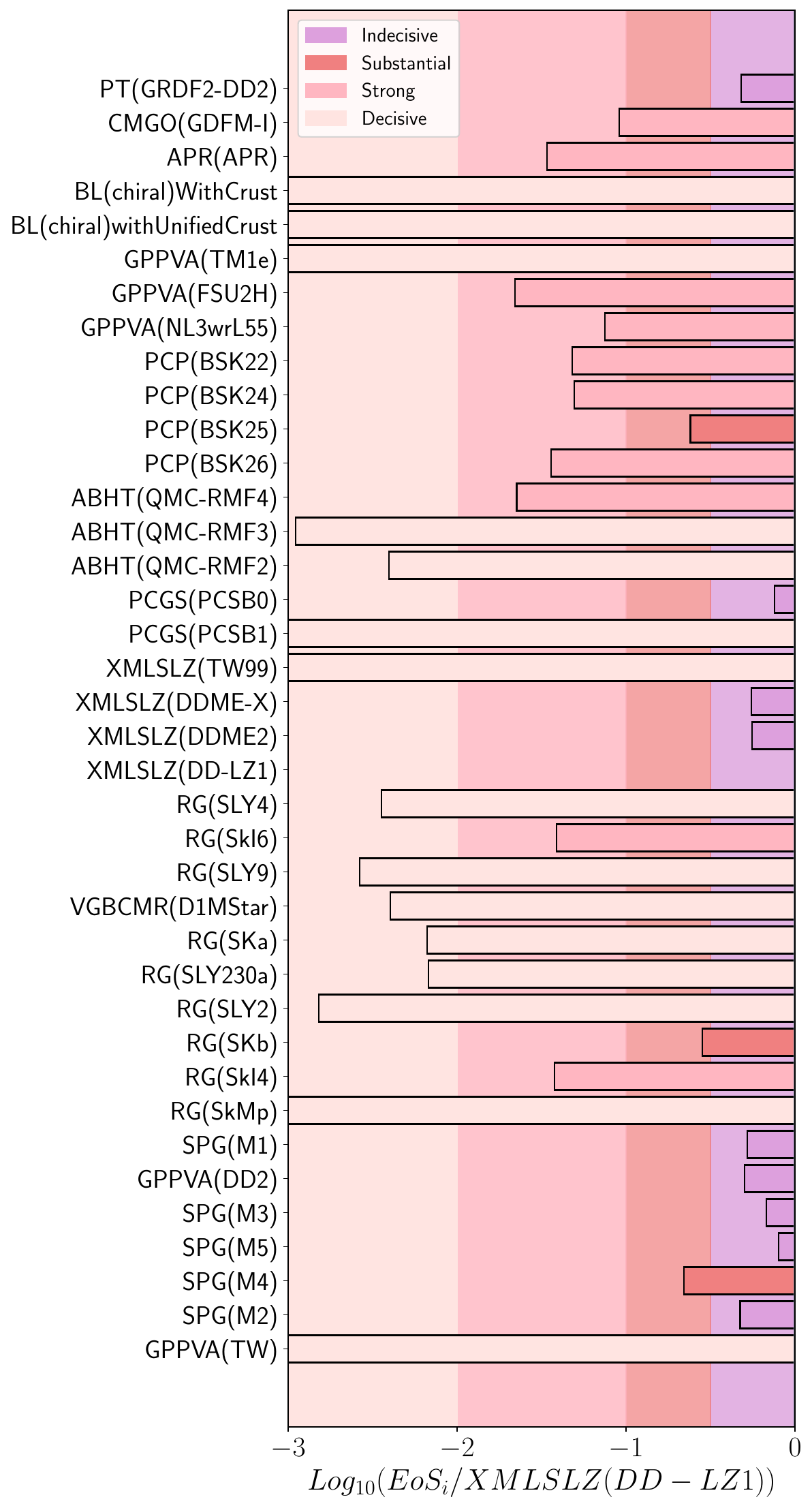}
    \caption{\raggedright \small Odds ratio plot of {XMLSLZ(DD-LZ1)} with other hadronic EoSs. Following Jeffrey's scale, the region between $(-0.5,0)$ (shaded with `plum') is the region in which, if a model lies, it provides evidence for the base model, but it is not worth more than a bare mention (indecisive). The region between $(-1,-0.5)$ (shaded with `light coral') is the region of substantial evidence for the base model, and the region between $(-2,-1)$ (shaded with `light pink') is the region of strong evidence for the base model. The region beyond $-2$ (shaded with `misty rose') is the region of decisive evidence for the base model. The histogram for the odds ratio value of an EoS is depicted in the corresponding colour of the respective region it lies in; for example, if an EoS is situated in the decisive region, its histogram is coloured in `misty rose'. $EoS_i$ refers to the equation of state being compared.} \label{fig:Oddsplothadronic}
\end{figure}

After evaluating the evidence value for each EoS, the EoS with the highest evidence value is {XMLSLZ(DD-LZ1)}, which uses the Thomas Fermi approach. Fig. \ref{fig:Oddsplothadronic} shows the odds ratio plot of each EoS with respect to {XMLSLZ(DD-LZ1)}.

Upon utilizing Jeffrey's scale, as discussed before, there are precisely {15} EoS that can be decisively rejected based on their odds ratio value with respect to {XMLSLZ(DD-LZ1)}. Table \ref{tab:had} highlights the rejected EoSs with bold text. Additionally, there are 9 other EoSs with odds ratio values situated in the `indecisive' region (\cref{fig:Oddsplothadronic}). They are based on the following EoS models: density-dependent RMF model, Thomas Fermi approximation, RMF model, and generalized relativistic density functional.

\begin{figure*}
    \centering
    \begin{subfigure}{0.45\textwidth}
        \centering
        \includegraphics[width=\linewidth]{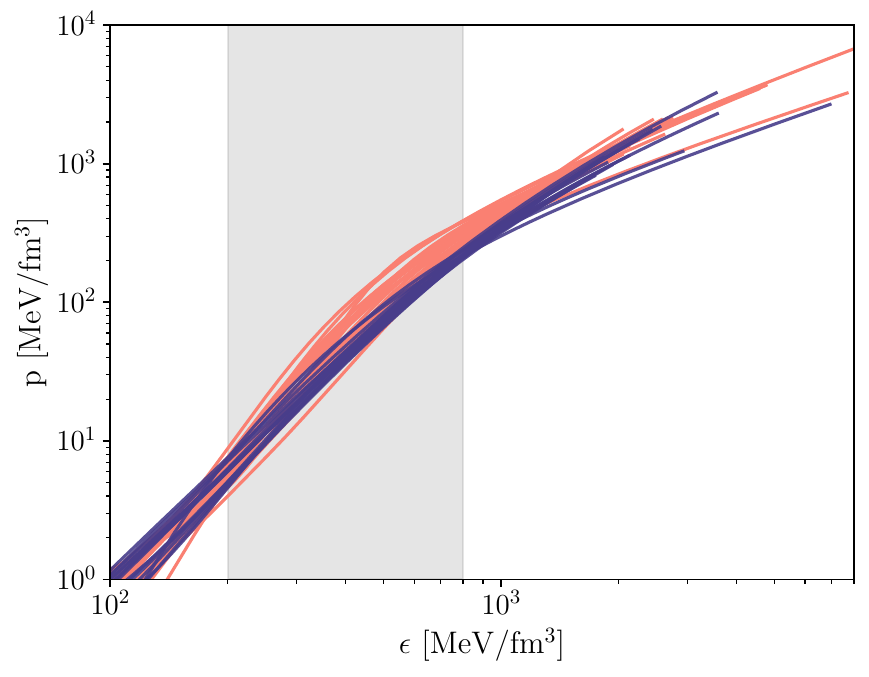} 
        \caption{}
        \label{fig:RA_Hadronic_EoS}
    \end{subfigure}
    \begin{subfigure}{0.45\textwidth}
        \centering
        \includegraphics[width=\linewidth]{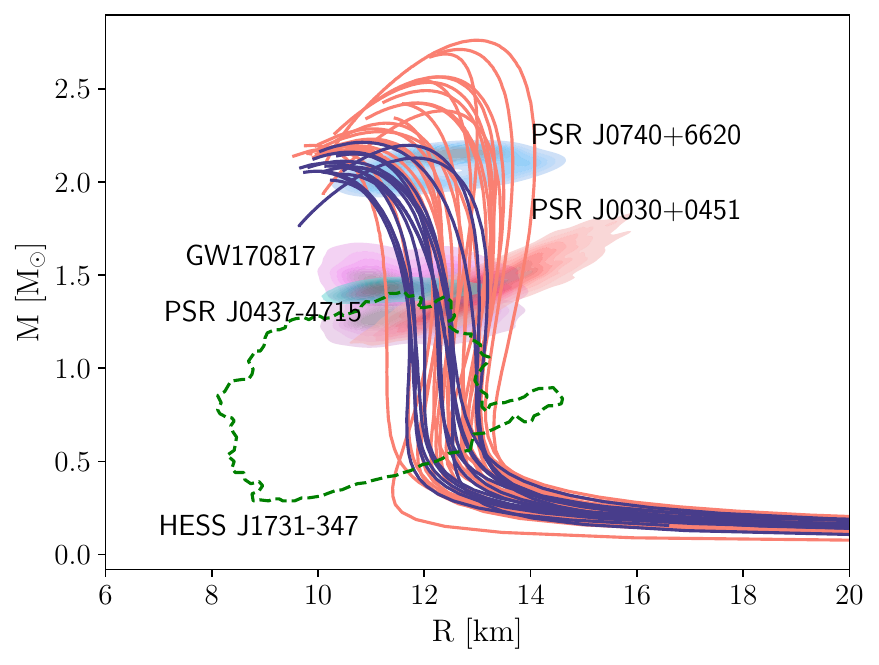} 
        \caption{}
        \label{fig:RA_Hadronic_MR}
    \end{subfigure}

    \begin{subfigure}{0.45\textwidth}
        \centering
        \includegraphics[width=\linewidth]{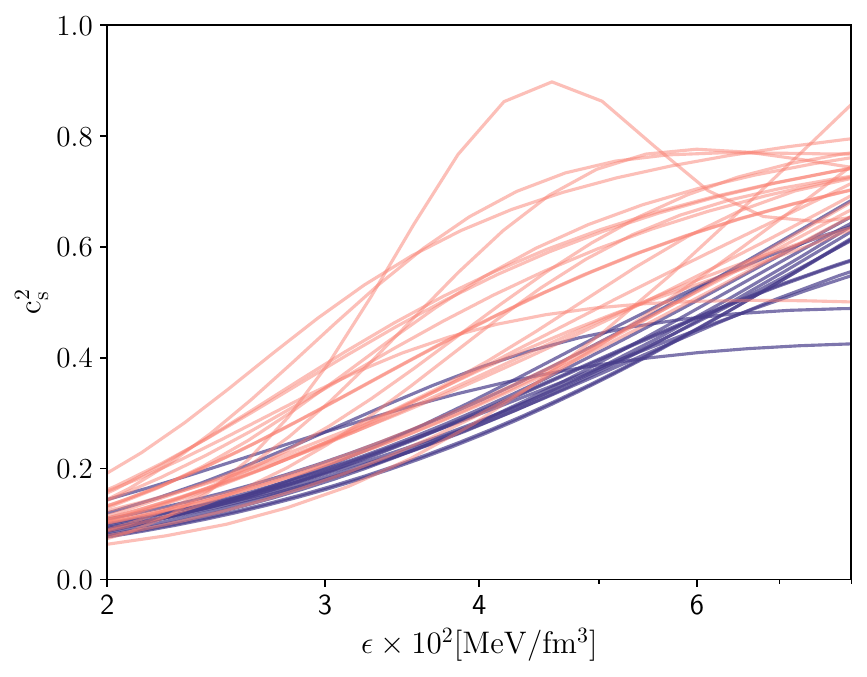} 
        \caption{}
        \label{fig:RA_Hadronic_Cs2}
    \end{subfigure}
    \begin{subfigure}{0.45\textwidth}
        \centering
        \includegraphics[width=\linewidth]{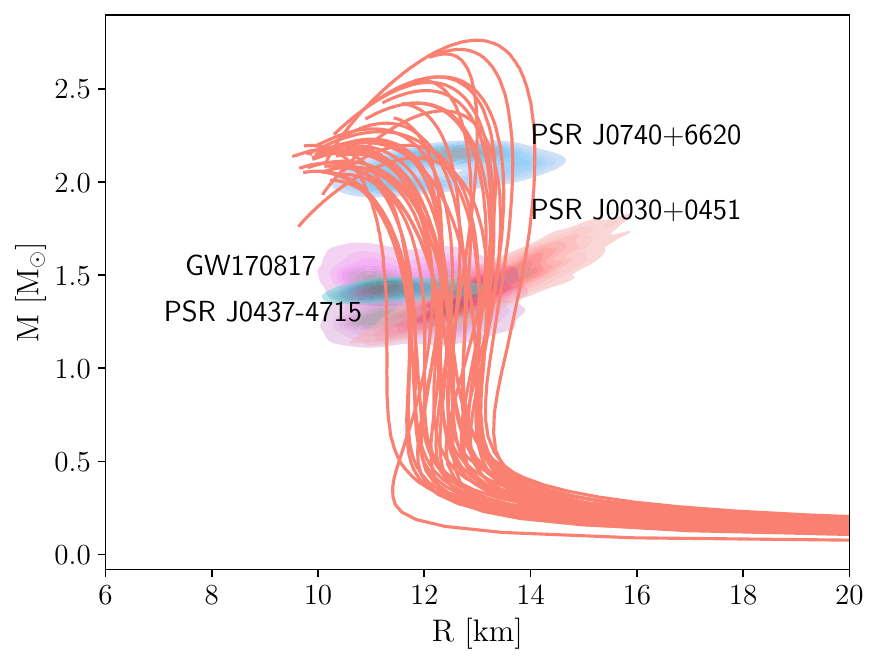} 
        \caption{}
        \label{fig:RA_Hadronic_onlypulsarMR}
    \end{subfigure}
    
    \caption{\raggedright  \textbf{(a):} EoS plot of the hadronic EoSs. The grey patch shows the region where there is a change in the stiffness of EoS; \textbf{(b):} Mass-Radius curves of the hadronic EoSs; \textbf{(c):} Illustrates the speed of sound corresponding to the shaded region in \textbf{(a)}; \textbf{(d):} Illustrates the rejected and accepted hadronic EoSs when HESS J1731-347 is not considered. In all the figures, rejected EoSs are plotted in dark blue, while the accepted ones are plotted in salmon.}
    \label{fig:Hadronic_Rejected}
\end{figure*}

Using \cref{tab:had} as a reference, it can be observed that except the EoSs based on Brussels-Montreal energy density functional, every other type of EoS has at least one EoS that is being rejected. However, none of the EoSs based on the Brussels-Montreal energy density functional lie in the `indecisive' region. 
On the contrary, while one of the EoS based on the density-dependent RMF model can be decisively rejected, 5 (out of 7) of the EoSs lie in the `indecisive' region. {Similarly, 3 (out of 4) of the EoSs following the Thomas Fermi approximation lie in the `indecisive' region, with XMLSLZ(DD-LZ1) as the best performing EoS}. 
This analysis suggests that the density-dependent RMF model {and the Thomas Fermi approximation} best satisfy all the current astrophysical observations. Additionally, the EoS PT(GRDF2-DD2) (based on generalized relativistic density functional) also performed exceptionally well.

Using \cref{tab:had} and \cref{fig:Oddsplothadronic}, it can be seen that the effective-interaction based model, although not decisively rejected, is the least effective model when explaining current astrophysical bounds. Furthermore, it should be noted that all EoSs based on the Brueckner-Hartree-Fock approximation were rejected, leading us to the conclusion that it is the least plausible EoS type that could explain the current observations, including HESS J1731-347.

{Most of the EoSs were accepted or rejected based on either the radius or the tidal deformability bound. Since the observations in consideration are rather recent, those EoSs that were old and did not maintain the radius or tidal deformability bounds were found to be rejected.
Additionally, using \cref{tab:had} and \cref{fig:Oddsplothadronic} as a reference, it can be seen that the comparable EoSs have very similar radii for R$_{1.4}$ and R$_{2.0}$. We notice a trend of increasing evidence against EoSs that have increasingly dissimilar values for R$_{1.4}$ and R$_{2.0}$. A similar trend was also observed by Ref. \citep{Xia_2022}.}

In \cref{fig:RA_Hadronic_EoS,fig:RA_Hadronic_MR}, we show the comparison of the EoSs that were rejected along with their MR curves. { Fig. \ref{fig:RA_Hadronic_MR} shows that the accepted EoSs have an MR curve that has a back-bending effect above 0.5 M$_{\odot}$. They also support a larger maximum mass. Additionally, the back-bending of the MR curve is reflected in the sudden stiffening of the EoS beyond $200$ MeVfm$^{-3}$ in \cref{fig:RA_Hadronic_EoS}. However, at much higher densities, the curves become softer. This shows that the best performing EoSs are more non-monotonous than the rejected ones.} 

{This non-monotonous nature can be examined in greater detail in the speed of sound plot.}
The adiabatic speed of sound ($c_s = \sqrt{\partial p / \partial \epsilon}$) is an essential quantity as it determines the slope of the EoS \citep{FERRER2023122608, Bedaque, Reed, Tews_2018,Chatterjee:2023ecc}. {Fig. \ref{fig:RA_Hadronic_Cs2} shows an interesting feature; the accepted curves are clearly more non-monotonic than the rejected curves, with a few even attaining a local maximum. Usually, this is associated with the production of certain new degrees of freedom or suppression of some existing degrees \citep{Zhao_Lattimer_cs2dof,Mclerran_Jeong_cs2dof,Mclerran_Reddy_cs2dof}.}

Fig. \ref{fig:RA_Hadronic_onlypulsarMR} shows the rejected and accepted mass-radius curves of the hadronic EoSs when the data of HESS J1731-347 was not considered. Contrary to the scenario when we consider the observation of HESS J1731-347, none of the hadronic EoSs used in this work could be decisively rejected in this scenario, as shown in \cref{fig:RA_Hadronic_onlypulsarMR}. The rejection of the EoSs from our analysis shows that all the current astrophysical observations prefer a stiffer hadronic EoS at intermediate densities.

\subsection{Strange EoS}

\begin{figure}
    \centering 
    \includegraphics[scale=0.46]{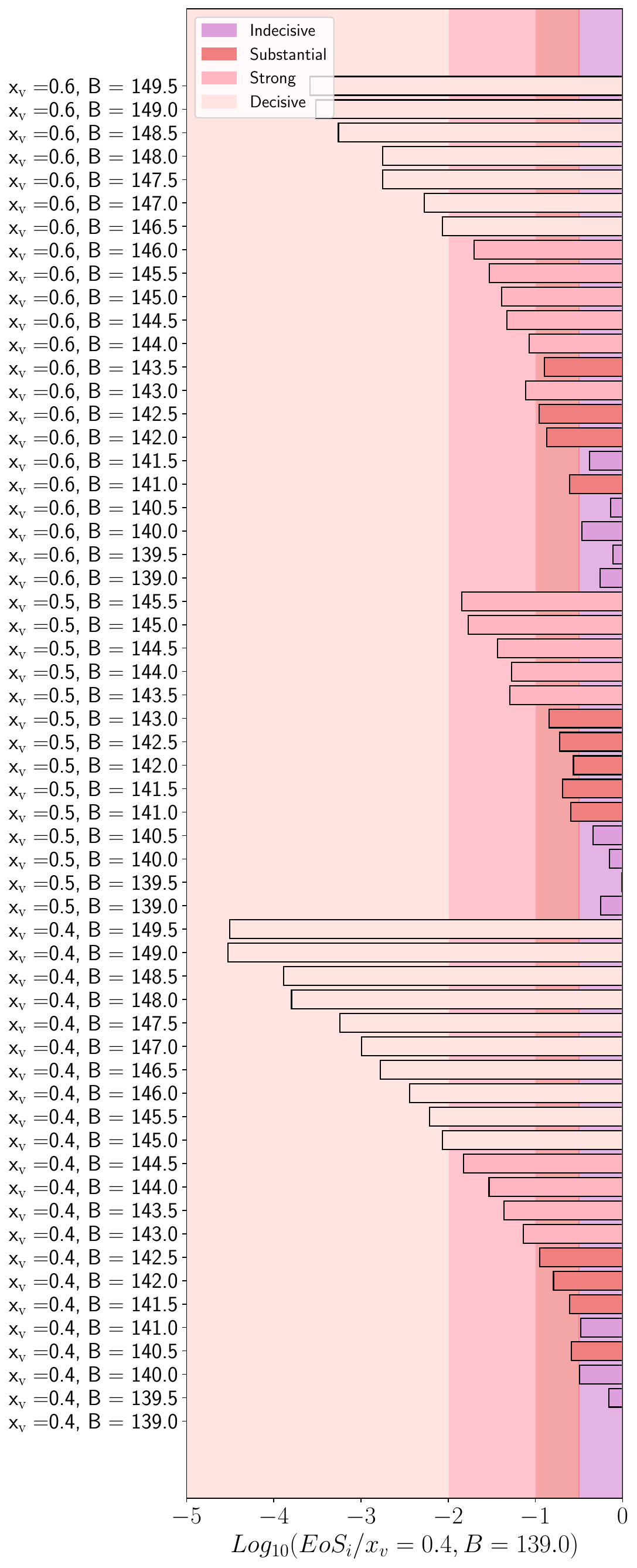}
    \caption{\raggedright  \small Odds ratio plot of the strange matter EoS with {$x_v = 0.4$ and B = 139.0} with other strange matter EoSs. The colour scheme is the same as in \cref{fig:Oddsplothadronic}. The ticks on the y-axis refer to strange matter EoSs with the corresponding $x_v$ and B parameters. $EoS_i$ refers to the equation of state being compared.} \label{fig:Strangeoddsplot}
\end{figure}

The strange matter EoSs were constructed by varying the scaled coupling constant  `$x_v$' and bag parameters `B' by small intervals as discussed in \cref{subsec:quark_eos}. Out of the 58 constructed EoSs, the EoS with {$x_v=0.4$ and B=139.0} possesses the highest evidence value.

\begin{figure*}
    \centering
    \begin{subfigure}{0.45\textwidth}
        \centering
        \includegraphics[width=\linewidth]{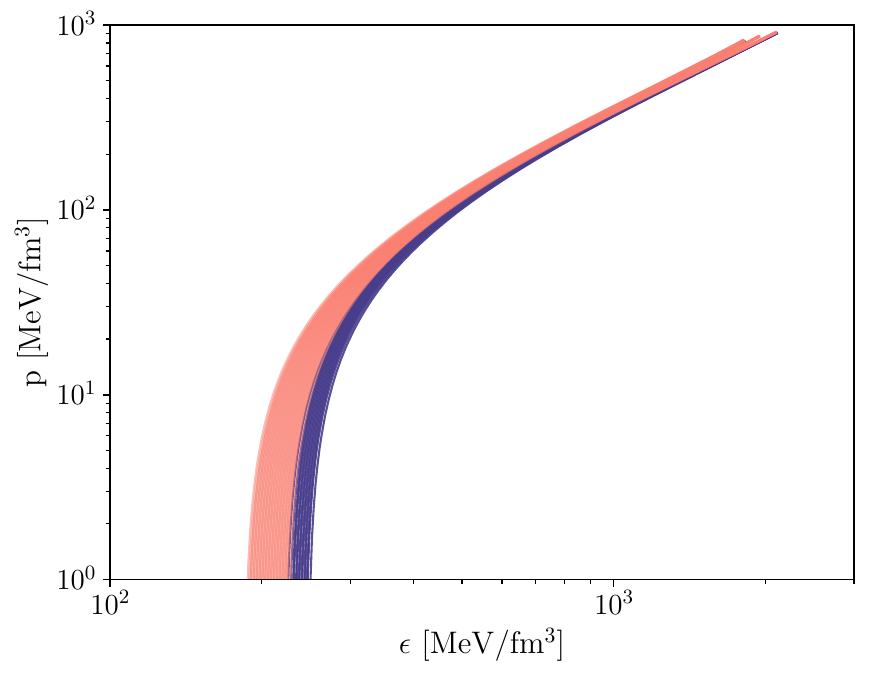} 
        \caption{}
        \label{fig:RA_Strange_EoS}
    \end{subfigure}
    \begin{subfigure}{0.45\textwidth}
        \centering
        \includegraphics[width=\linewidth]{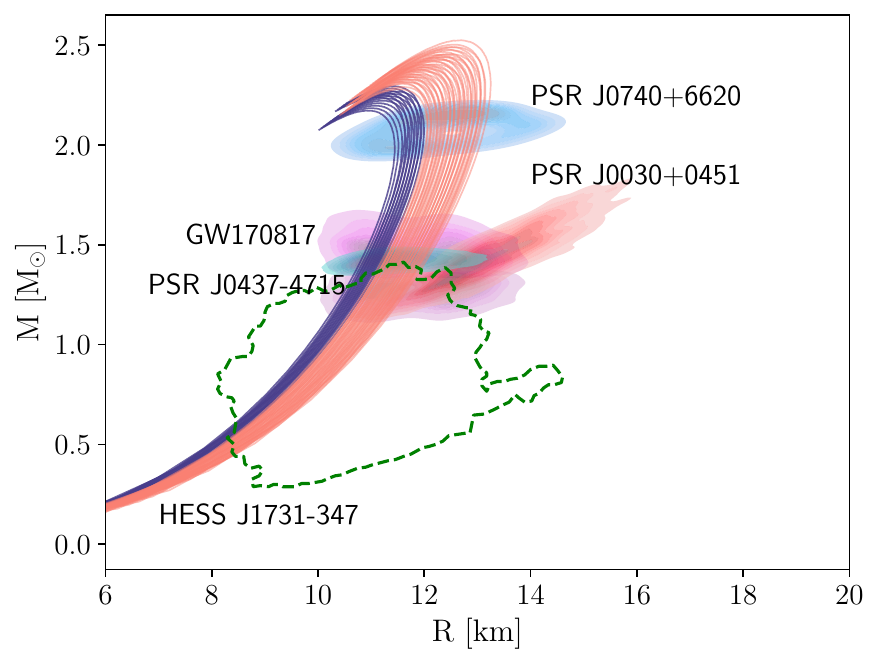} 
        \caption{}
        \label{fig:RA_Strange_MR}
    \end{subfigure}
    \begin{subfigure}{0.45\textwidth}
        \centering
        \includegraphics[width=\linewidth]{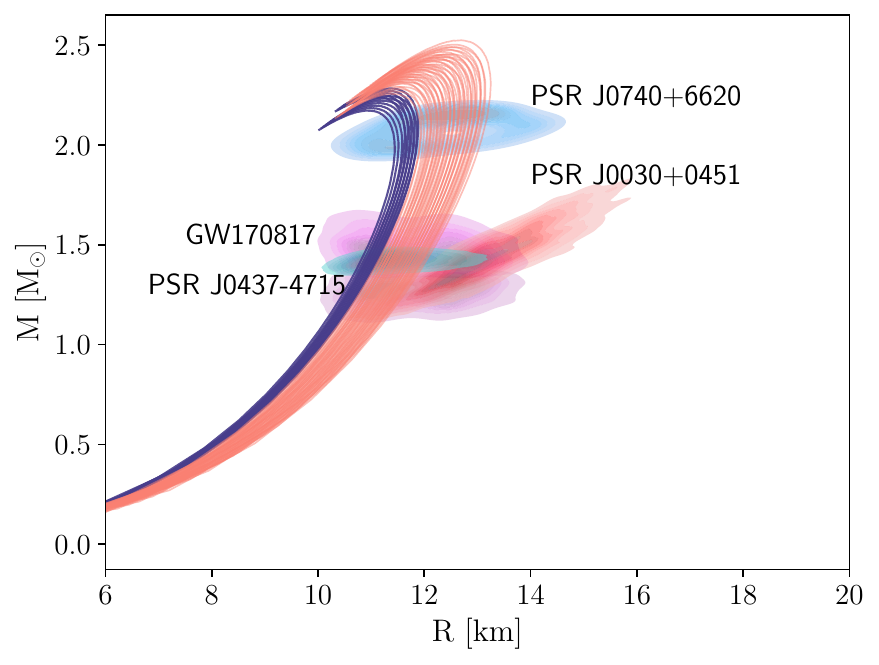} 
        \caption{}
        \label{fig:RA_Strange_onlypulsarMR}
    \end{subfigure}
    
    \caption{\raggedright \textbf{(a):} EoS plot of the strange matter EoSs; \textbf{(b):} Mass-Radius curves of the strange matter EoSs; \textbf{(c):} Mass-Radius curves of the strange matter EoSs when HESS J1731-347 is not considered. The colour nomenclature remains the same as the previous one.}
    \label{fig:StrangeEoS}
\end{figure*}

Fig. \ref{fig:Strangeoddsplot} shows the odds ratio plot of each quark matter EoS in comparison to the EoS with {$x_v = 0.4$ and B = 139.0}. There are {12 other} EoSs situated in the `indecisive' region, all associated with low values of the bag parameter. Fig. \ref{fig:Strangeoddsplot} shows a clear trend of increasing evidence against EoSs with increasing values of the bag parameter, {with only two minor deviations from it. This should be attributed to the fact that increasing the value of the bag parameter decreases the stiffness of the EoS. Therefore, our analysis suggests that similar to the hadronic matter EoSs, strange matter EoSs also prefer stiff EoSs.}

We found that {17} EoSs could be decisively rejected, whose bag parameter values were found to be $\ge$ 146.5. Furthermore, for $x_v=0.4$, {10} EoSs (with B $\ge$ 145.0) were found to lie in the 'decisive' region, whereas, for $x_v=0.6$, 7 EoSs (with B $\ge$ 146.5) were found to lie in the `decisive' region, implying that by increasing the value of the scaled coupling constant, some higher values of the bag parameter could still be preferred by the current observations. {This is an expected result since an increase in the value of the scaled coupling constant increases the stiffness of the EoS.}

The rejected and accepted EoSs and their corresponding M-R curves are shown in \cref{fig:RA_Strange_EoS,fig:RA_Strange_MR} respectively. {In the M-R plot, the EoSs that barely cut the PSR J0030+0451 contour are rejected, and those that have a significant overlap with the contour are accepted by our analysis. Looking at \cref{fig:RA_Strange_EoS}, it can be seen that at low pressure, there is a cutoff value of density for the EoS to be rejected (approximately 225 MeVfm$^{-3}$).
}

From \cref{fig:RA_Strange_MR,fig:RA_Strange_onlypulsarMR}, we see the difference in MR curves when we do not include HESS J1731-347 data. Without the inclusion of this observation, only a few EoSs were found to be decisively rejected; however, this number significantly increased upon its inclusion.


\subsection{Hybrid EoS}

\begin{figure}[H]
    \centering
    \begin{subfigure}{0.55\textwidth}
        \centering
        \includegraphics[width=\textwidth]{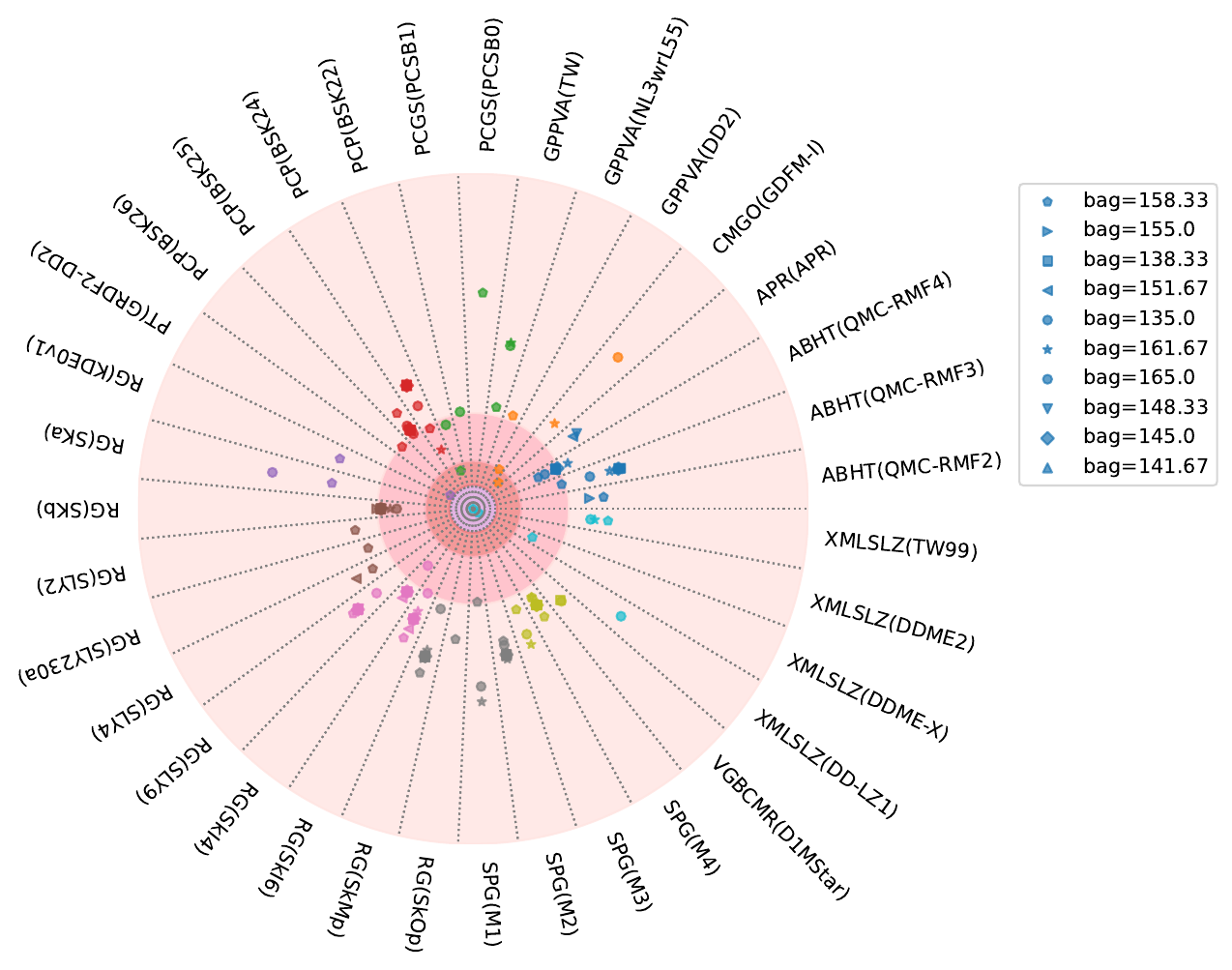}
        \caption{}
        \label{fig:Hybrid_Odds_xv044}
      
    \end{subfigure}
   \hfill
   \begin{subfigure}{0.55\textwidth}
        \flushleft
        \includegraphics[width=0.82\textwidth]{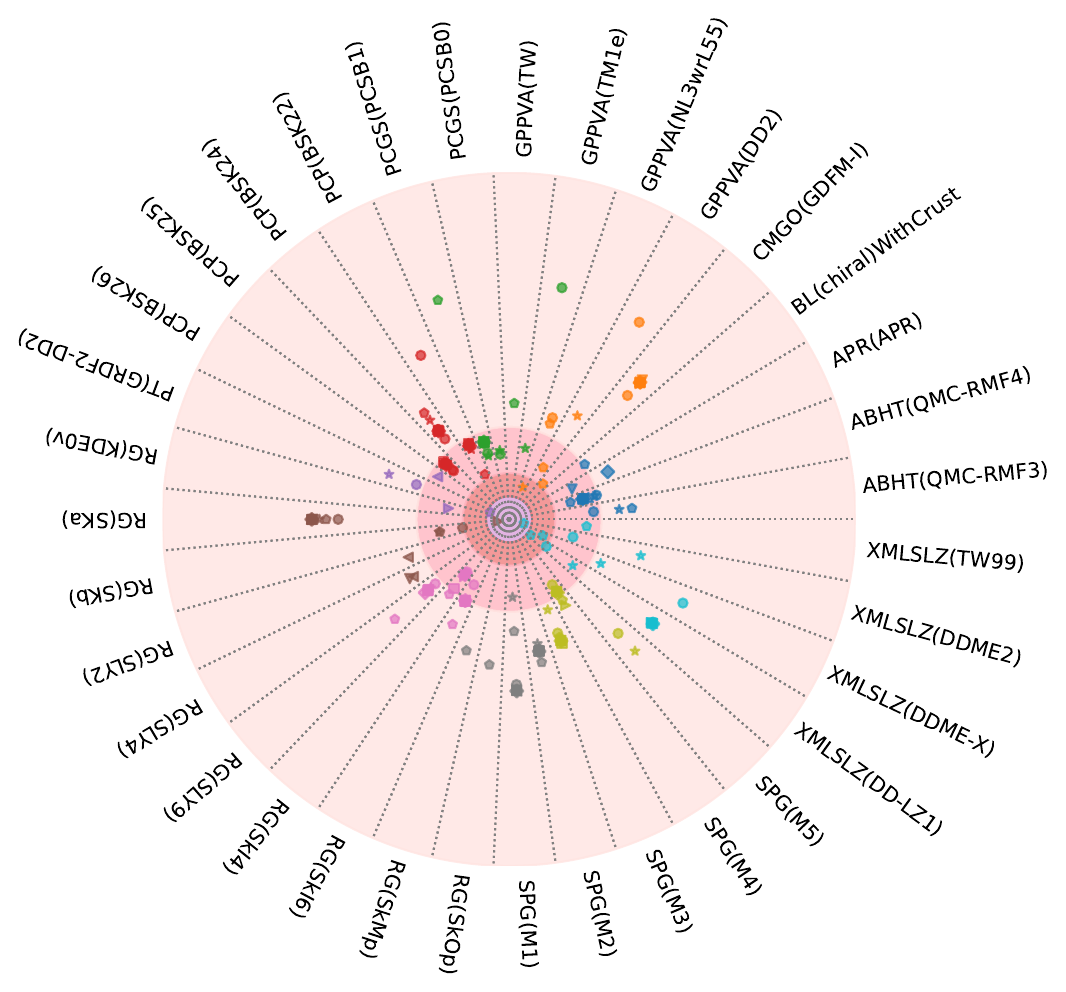}
        \caption{}
        \label{fig:Hybrid_Odds_xv05}
       
    \end{subfigure}
   \hfill
    \caption{\raggedright \textbf{(a)}: Odds ratio plot of Hyb\_best with EoSs having a fixed $x_v$ value of 0.44; \textbf{(b)}: Odds ratio plot of Hyb\_best with EoSs having a fixed $x_v$ value of 0.5. The colour nomenclature of the disks is the same as in \cref{fig:Oddsplothadronic,fig:Strangeoddsplot}. Each sector belongs to a hadronic family. The different shaped points correspond to different bag values referenced in \textbf{(a)}. The different colours of the points help distinguish continuous sets of hadronic families. The odds ratio value of each EoS acts as the radius value in the plot. This value is then plotted in the sector of the hadronic family of that EoS, with the shape of its corresponding bag value.}
    \label{fig:Hybrid_xv044_xv05}
\end{figure}

\begin{figure}[H]
    \flushright
    \includegraphics[width=0.9\linewidth]{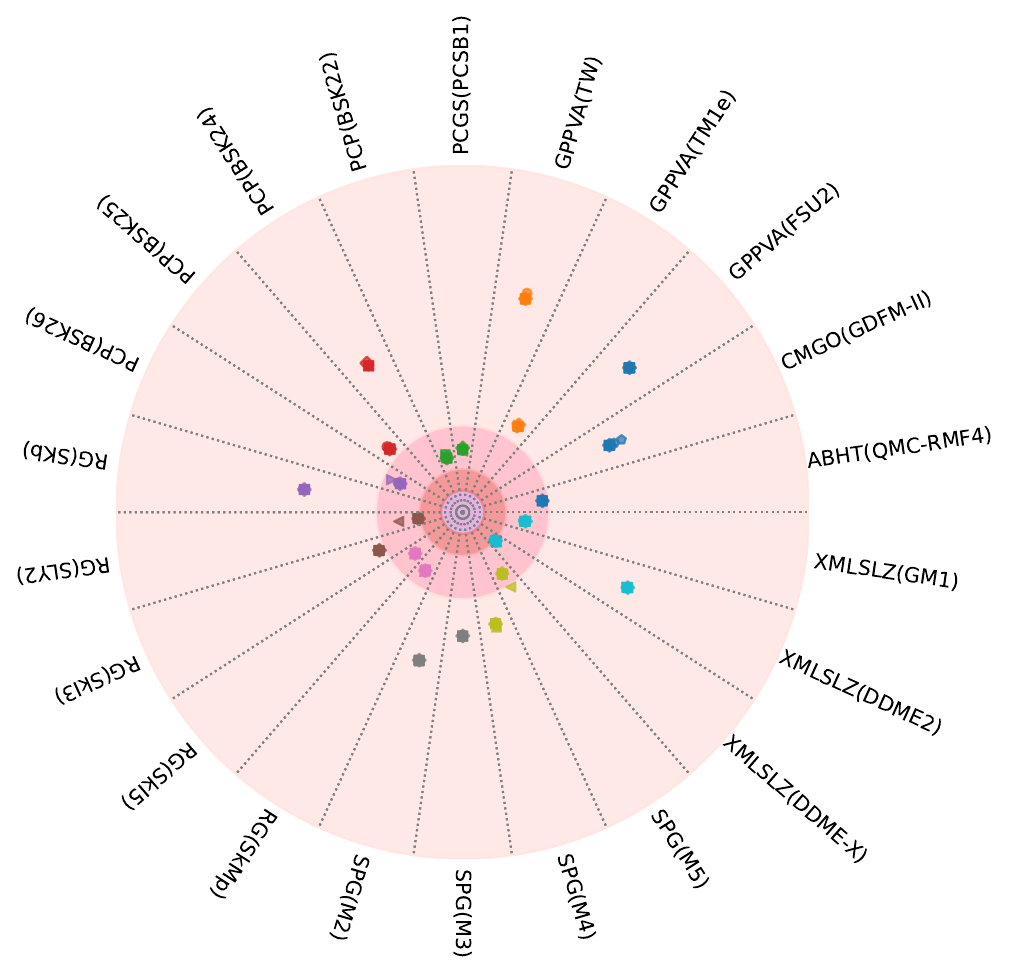}
    \caption{\raggedright Odds ratio plot of Hyb\_best with EoSs having a fixed $x_v$ value of 0.2. The nomenclature is the same as \cref{fig:Hybrid_xv044_xv05}.}
    \label{fig:Hybrid_xv02}
\end{figure}

The hybrid EoSs constructed ({545} are used in our analysis) in \cref{subsec:PT_EoS} were evaluated. {XMLSLZ-DDME2 with $x_v = 0.44$ and B = 158.33} was found to have the highest evidence value. In the following sections, we shall denote this EoS as `Hyb\_best' to avoid lengthy phrases. Figs. \ref{fig:Hybrid_xv044_xv05} and \ref{fig:Hybrid_xv02} show the odds ratio plot of each EoS (for different $x_v$ values) with respect to Hyb\_best.

A total of {351} EoSs were `decisively' rejected when compared with Hyb\_best, and there are {4 other} EoSs with odds ratio values situated in the `indecisive' region. The comparable EoSs are: {XMLSLZ-DDME2 ($x_v = 0.5$, B=158.33), XMLSLZ-DDLZ1 ($x_v = 0.44$, B=158.33), RG-SKb ($x_v = 0.5$, B=155.0) and PT-GRDF2-DD2 ($x_v = 0.5$, B=158.33)}. Additionally, it is observed that among the {five} comparable EoSs, including Hyb\_best, {four} of them exhibit a bag parameter value of 158.33, indicating a greater preference for this specific value. However, it should be noted that such a preference is only observed in higher values of the scaled coupling constant (\cref{fig:Hybrid_xv044_xv05}) and not in lower values. {It is also evident in \cref{fig:Hybrid_xv02} that choosing B=158.33 for $x_v = 0.2$ does not improve the odds ratio value of the EoS family.}

\begin{figure*}[!p]
    \centering
    \begin{subfigure}{0.4\textwidth}
        \centering
        \includegraphics[width=\linewidth]{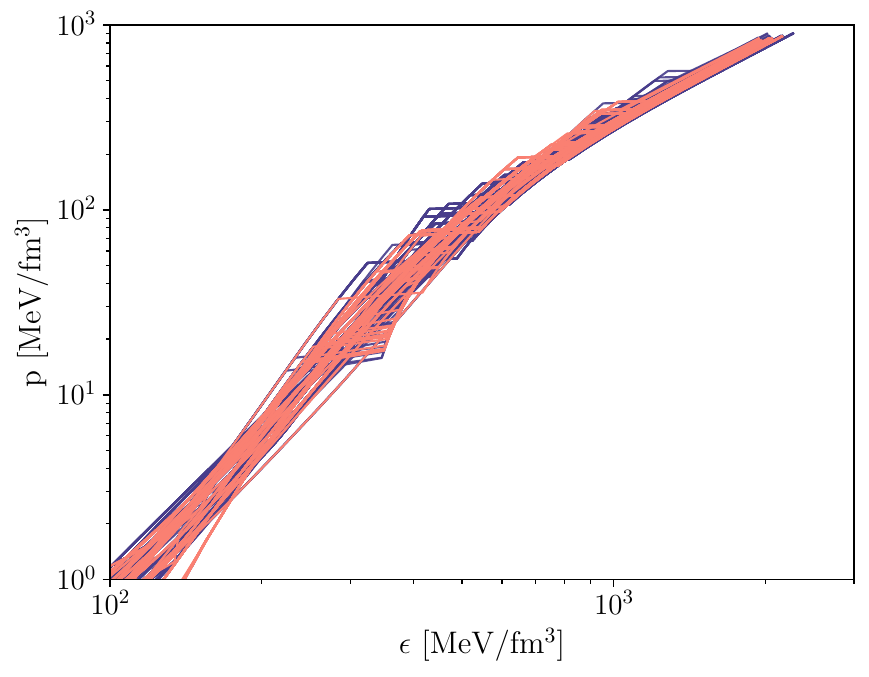} 
        \caption{}
        \label{fig:RA_Hybrid_EoS}
    \end{subfigure}
    \begin{subfigure}{0.4\textwidth}
        \centering
        \includegraphics[width=\linewidth]{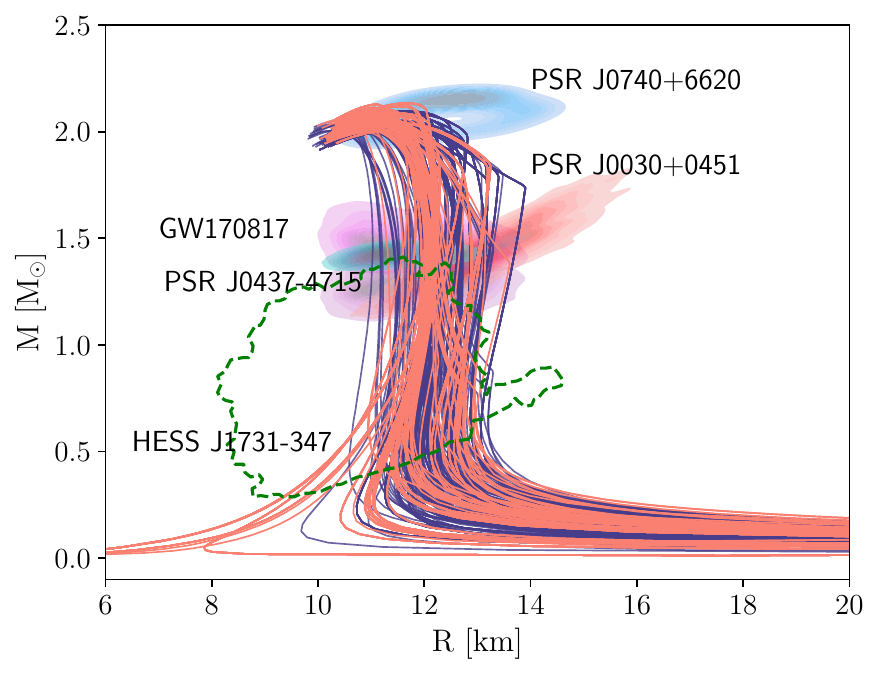} 
        \caption{}
        \label{fig:RA_Hybrid_MR}
    \end{subfigure}
    \begin{subfigure}{0.4\textwidth}
        \centering
        \includegraphics[width=\linewidth]{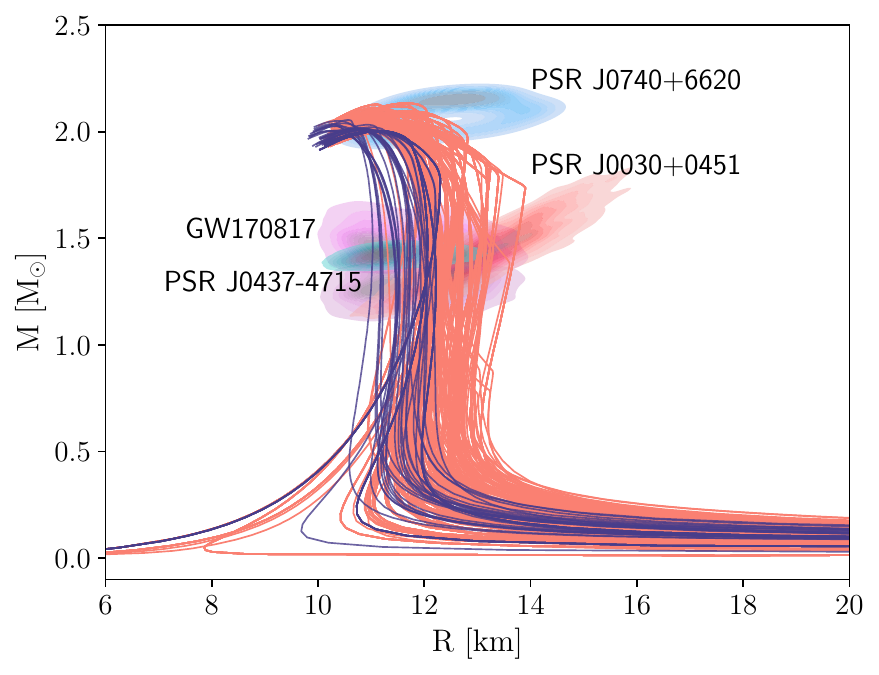} 
        \caption{}
        \label{fig:RA_Hybrid_onlypulsarMR}
    \end{subfigure}
    
    \caption{\raggedright \textbf{(a):} EoS plot of the hybrid EoSs; \textbf{(b):} Mass-Radius curves of the hybrid EoSs; \textbf{(c):} Mass-Radius curves of the hybrid EoSs when HESS J1731-347 is excluded. The colour nomenclature remains the same as the previous one.}
    \label{fig:HybridEoS}
\end{figure*}

\begin{figure*}[!p]
    \centering
    \begin{subfigure}{0.45\textwidth}
        \centering
        \includegraphics[width=\textwidth]{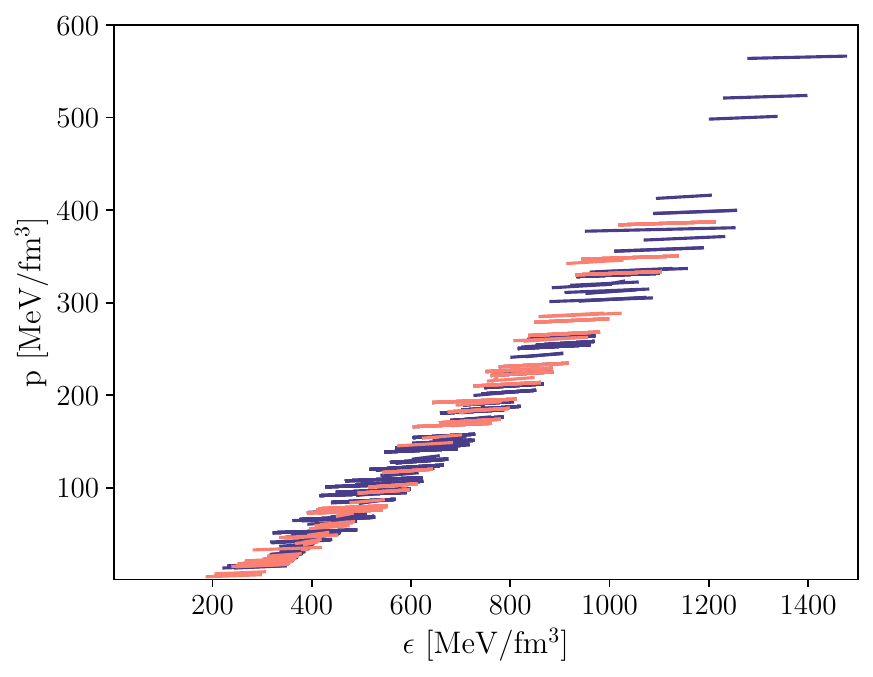}
        \caption{}
        \label{fig:Hybrid_Discontinuity}
    \end{subfigure}
   \hfill
   \begin{subfigure}{0.45\textwidth}
        \centering
        \includegraphics[width=\textwidth]{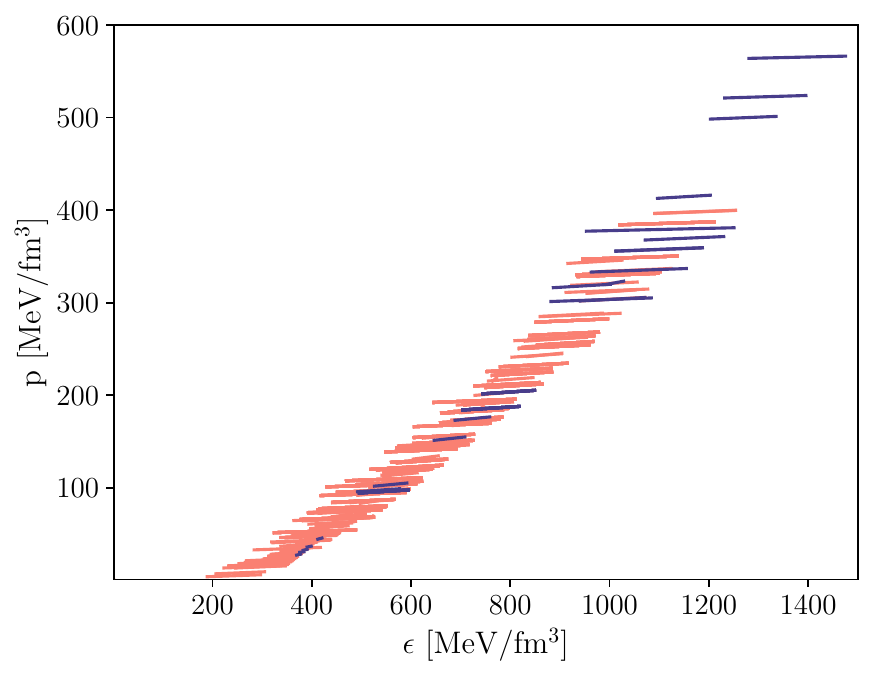}
        \caption{}
        \label{fig:Hybrid_Discontinuity_onlypulsar}
    \end{subfigure}
   \hfill
    \caption{\raggedright \textbf{(a)}: The density discontinuity corresponding to the hybrid EoSs from \cref{fig:RA_Hybrid_MR} is shown; \textbf{(b)}: The density discontinuity corresponding to the hybrid EoSs from \cref{fig:RA_Hybrid_onlypulsarMR} is shown. The colour nomenclature remains the same as the previous one.}
    \label{fig:Discontinuity_plot}
\end{figure*}

In \cref{fig:Hybrid_xv044_xv05,fig:Hybrid_xv02}, certain EoS families do not exhibit much change in their evidence values, even after changing the scaled coupling constant and bag parameter values. Figs. \ref{fig:RA_Hybrid_EoS} and \ref{fig:RA_Hybrid_MR} show the accepted and rejected hybrid EoSs and their corresponding M-R curves, respectively. {Since it is difficult to comment on the hybrid EoSs using only \cref{fig:RA_Hybrid_MR}, we utilize the two parameters of Maxwell's construction: width of the discontinuity and the onset of phase transition. 

Fig. \ref{fig:Hybrid_Discontinuity} illustrates both these parameters for the rejected and accepted EoSs. It can be seen that EoSs with an early onset of PT are preferred over those with a PT at higher densities. We conclude that the onset density of PT is more important, and the preference of the EoS does not depend on the width of the discontinuity.}

Fig. \ref{fig:RA_Hybrid_onlypulsarMR} illustrates the M-R curves of hybrid EoSs when the data on HESS J1731-347 is not considered. Comparing \cref{fig:RA_Hybrid_MR} and \cref{fig:RA_Hybrid_onlypulsarMR}, we find that the inclusion of the information of HESS J1731-347 dramatically increases the number of EoSs being rejected and puts a better constraint on the nature of the EoS. Without HESS J1731-347 information, only EoSs having higher radii were favoured. Including the data of HESS J1731-347, stars with smaller radii cannot be rejected from our analysis. Analyzing \cref{fig:Hybrid_Discontinuity_onlypulsar} indicates that no comments can be made upon the transition of FOPT without the HESS data. However, upon including it (\cref{fig:Hybrid_Discontinuity}), one can definitely say that EoSs with early FOPTs are preferred by the observations.

\subsection{Comparison among the `Indecisive' EoSs}

\begin{figure}
    \centering
    \includegraphics[width=0.48\textwidth]{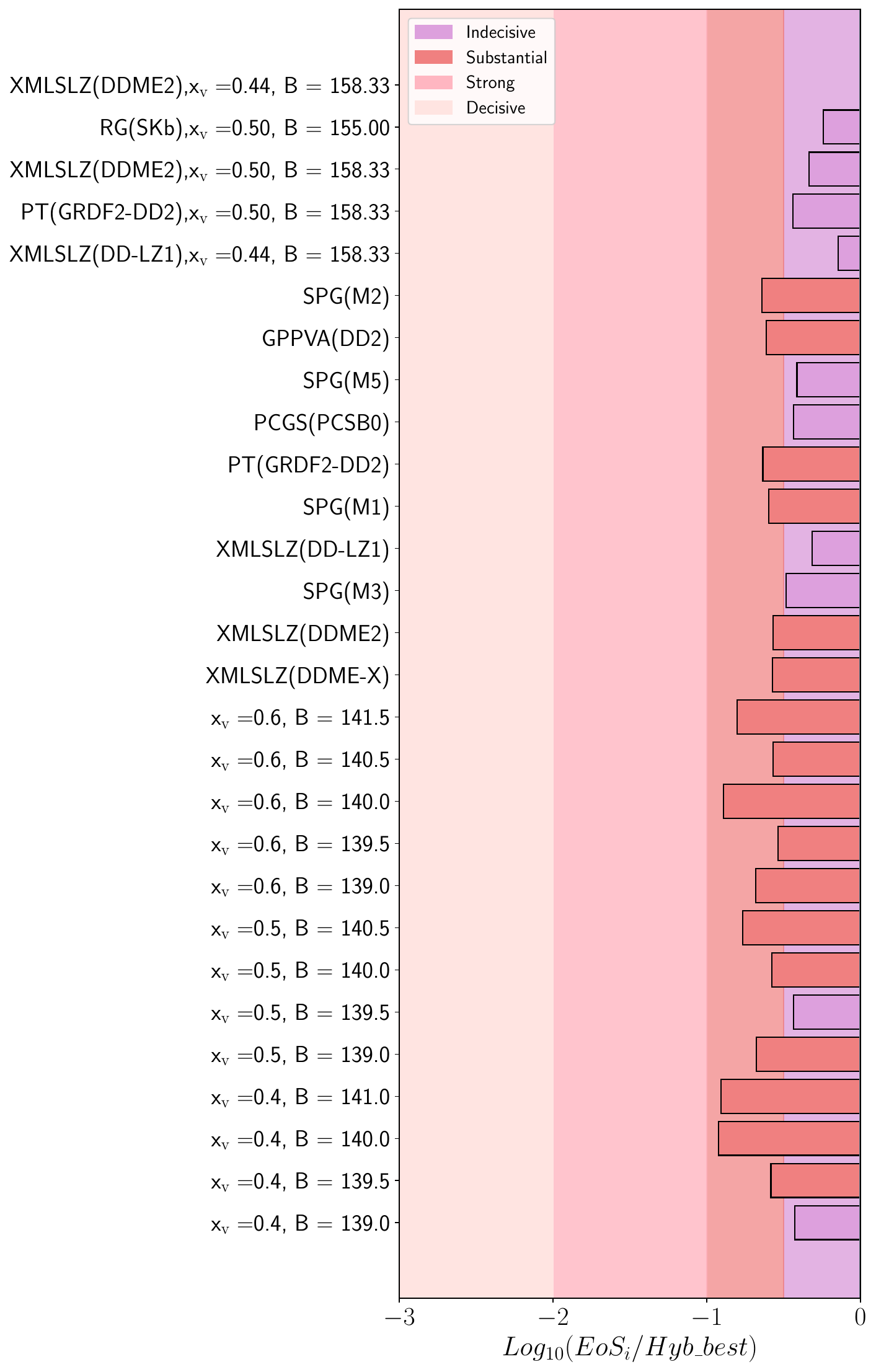}
    \caption{\raggedright Odds ratio plot of Hyb\_best with the best performing EoSs in each family of EoSs. Note that EoS above refers to Hyb\_best, and $EoS_i$ refers to the equation of state being compared.}
    \label{fig:Compare}
\end{figure}

In the previous subsections, we have found the best performing EoS from the hadronic family, strange matter family, and the hybrid star family of EoSs. In this section, the odds ratio analysis among the EoSs from each family that lie in the `indecisive' region is performed. Fig. \ref{fig:Compare} shows the odds ratio plot of each EoS with Hyb\_best. It shows that the hybrid EoS Hyb\_best performs best among all the EoSs. We find that for the hybrid EoSs, no EoSs lie beyond the `indecisive' region, suggesting that the hybrid EoSs are the most probable among all the families of EoSs. The analysis of the hadronic EoSs reveals that EoSs following the RMF model, the Density-Dependent RMF model, and the Thomas-Fermi approach satisfy the current astrophysical observations the best. For the SM EoSs, we see that the EoSs having a value of B $\leq 139.5$ are most preferred.

\newpage

\section{Summary and Conclusions} \label{sec:Summ}
\vspace{-5pt}

The work improves the current status of the EoSs upon including the mass and radius measurements of the compact object HESS J1731-347 using a Bayesian model selection technique. The current astrophysical observations of - PSR J0030+0451, PSR J0740+6620, {J0437-4715}, and GW170817, along with HESS J1731-347, were used to unravel the implications of the latter observation. Starting with constructing three families of EoSs - the first family consists of hadronic matter. The second family of EoSs was built on the modified MIT bag model with scaled couplings and comprising of three flavoured quarks. For the third family of EoSs, we used hybrid EoSs comprising a FOPT.

After obtaining the EoSs, a Bayesian model selection was performed for each family of EoSs. In order to evaluate the odds ratio of the EoSs, the evidence integrals were solved by choosing mass as the parameter to be integrated. A uniform prior on mass was adopted, ranging from $0.5 M_{\odot}$ to the maximum mass allowed by the EoS. While calculating the odds ratio, we assume each EoS is equally likely, thus reducing the odds ratio to the Bayes factor. We adopted Jeffrey's scale, allowing us to decisively reject EoSs and choose the most suitable EoS models.

The analysis of the hadronic family of EoSs shows that the nuclear EoSs following the Brueckner-Hartree-Fock approximation and EoSs based on effective interactions are least effective in explaining the current astrophysical observations along with the compact object HESS J1731-347. {XMLSLZ(DD-LZ1), following the Thomas Fermi approximation, was found to be the best performing EoS. Density-dependent RMF models and the Thomas Fermi approximation are the most effective models in explaining the current observations.}   

{Most of the EoSs were accepted or rejected based on either the radius or the tidal deformability bound of the observations. The accepted EoSs showed a back-bending effect on the MR curve, which is due to a sudden stiffening of the EoS. Moreover, these models showed a non-monotonous nature in their speed of sound, with few showing maxima, which can be associated with the emergence of new degrees of freedom at higher densities.}

Analyzing the SM EoS, it was found that an increase in the Bag value fails to satisfy the current data as it effectively makes the EoS soft. {Our analysis of the SM EoS shows that if current observations are SSs, they would inherently prefer an EoS with smaller values of the bag parameter. However, if we increase the value of the scaled coupling constant, slightly higher bag values can still be preferred. There is a clear distinction between the rejected and accepted SM EoSs, both in the EoS and MR plot, with the data of the pulsar PSR J0030+0451 playing a decisive role in the segregation of the SM EoSs. Additionally, the analysis hints towards an energy density cut-off value at low pressures. If the energy density of an EoS at low pressure lies above this threshold value, it is rejected by our analysis.}

The analysis of the hybrid EoSs shows that a hybrid EoS referred to as Hyb\_best is the most likely hybrid EoS having a bag parameter value equal to 158.33 and scaled coupling constant value equal to 0.44. It was found that {4} other EoSs are comparable to Hyb\_best, {three} of which also have bag parameter values equal to 158.33, indicating a greater preference towards this bag value. However, such a preference is only observed in higher values of the scaled coupling constant $x_v$. {Although the MR curve does help in distinguishing the accepted EoSs from the rejected EoSs, this segregation is more dependent on the onset density of phase transition, preferring early PT densities.}

On comparing all the EoSs of different families that lie in the `indecisive' region of our analysis, we find that all the hybrid EoSs perform the best, suggesting that explaining current astrophysical observations using a hybrid EoS is the most likely scenario.  

{The results are consistent with the recent observation that EoSs show a non-monotonic speed of sound. The non-monotonicity is usually associated with the appearance of extra degrees of freedom (in this case, quark matter), suggesting a phase transition at intermediate densities in neutron stars. This is reflected in Hyb\_best being the most favored EoS.}

It is to be noted that we have considered only the modified MIT Bag model with vector interactions as the quark counterpart in our hybrid EoSs, hence there lies a possibility that the rejected hybrid EoSs might be able to explain the current observations upon considering different types of quark counterparts. {Since the current analysis does not consider the Gibbs construction, there also lies a possibility of explaining the present astrophysical observations using a FOPT constructed using the Gibbs construction. All such extensions are our future endeavors.
}


\section*{Acknowledgement}
\vspace{-5pt}
The authors would also like to thank IISER Bhopal for providing the infrastructure for this work. The authors also acknowledge Prasanta Char for their helpful discussions. DK acknowledges Bhaskar Biswas for email correspondence. ST would like to thank Akshat Singh, Pratik Thakur, Shamim Haque, and Rishi Gupta for their helpful discussions. SC acknowledges the Prime Minister's Research Fellowship (PMRF), Ministry of Education Govt. of India, for a graduate fellowship. RM and DK acknowledge the Science and Engineering Research Board (SERB), Govt. of India, for monetary support through a Core Research Grant (CRG/2022/000663).


\section*{Data Availability}
\vspace{-5pt}
The data can be made available upon reasonable request by the authors.


\appendix
\titleformat{\subsection}
  {\normalfont\small\bfseries}     
  {Appendix~\Alph{subsection}:}    
  {1em}                            
  {\centering}      


\vspace{-6pt}

\subsection{Effect of the CET Constraint}\label{appx:noCET}
\vspace{-5pt}

{In \cref{sec:formalism}, we discussed the construction of the CET band in order to use it to filter EoSs. This section discusses the changes in our analysis when the CET constraint is not considered. Initially, our analysis included 50 hadronic EoSs and 637 hybrid EoSs.
We found that from the 50 hadronic EoSs considered, 12 of them did not satisfy the CET bound.

On performing the entire analysis with these EoSs, we saw that 8 out of the 12 EoSs that did not satisfy the CET constraint were already rejected from our analysis. Of the remaining 4 EoSs, two were in the `substantial' regime, and two were in the `strong' regime. Similarly, on performing the analysis for the initial 637 hybrid EoSs, we found that 93 EoSs did not satisfy the CET constraint. Out of these 92 EoSs, 86 were rejected from our analysis, 2 were in the `strong' regime, 3 were in the `substantial' regime, and 1 was found to be in the `indecisive' regime.}

\subsection{Status of EoSs without HESS J1731-347} \label{appx:noHess}

\begin{figure*}
    \centering
    \begin{subfigure}{0.54\textwidth}
        \centering
        \includegraphics[width=\textwidth]{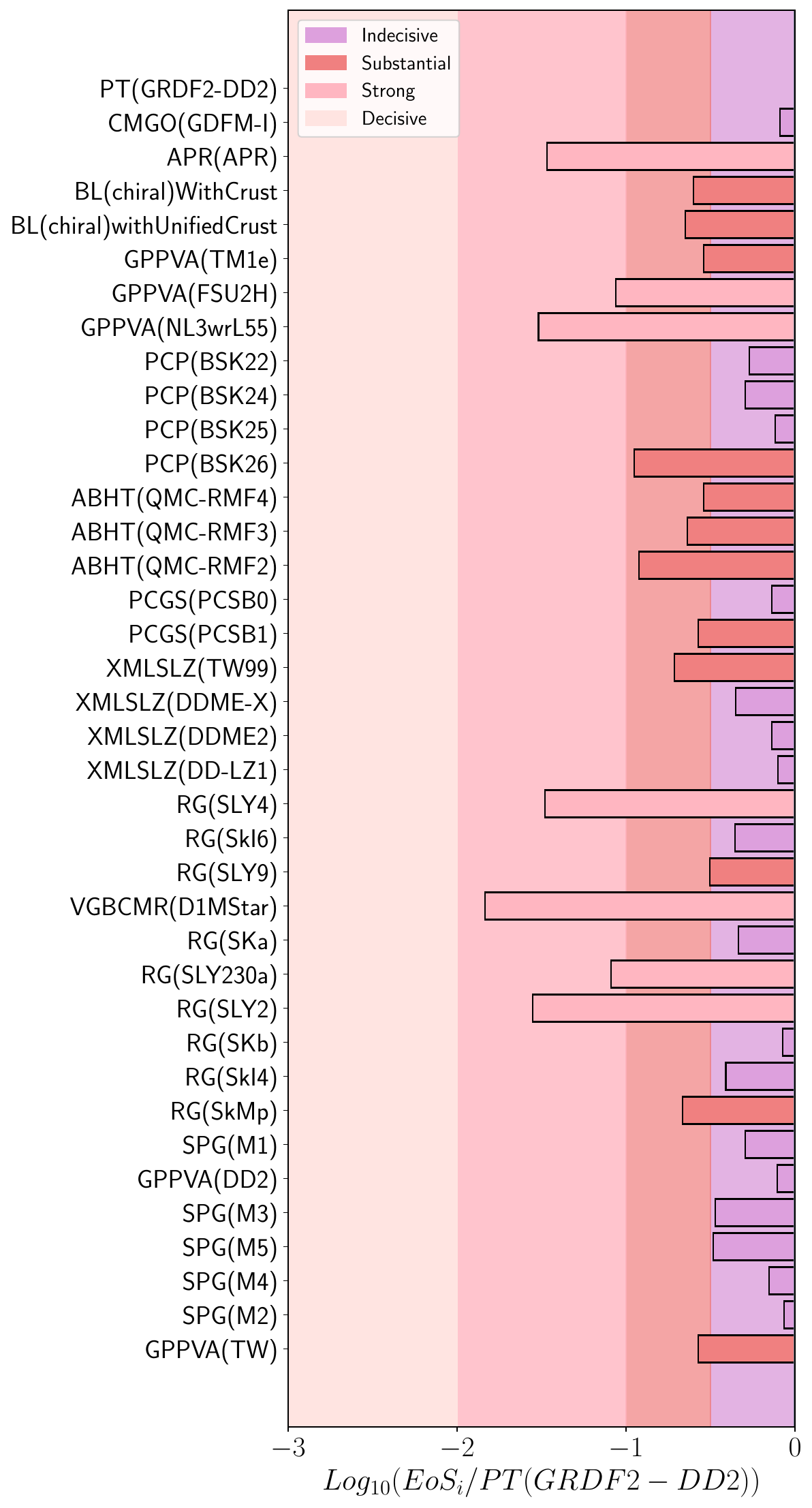}
        \caption{}
        \label{fig:Hadronic_Odds_onlypulsar}
    \end{subfigure}
    \begin{subfigure}{0.45\textwidth}
        \centering
        \includegraphics[width=\textwidth]{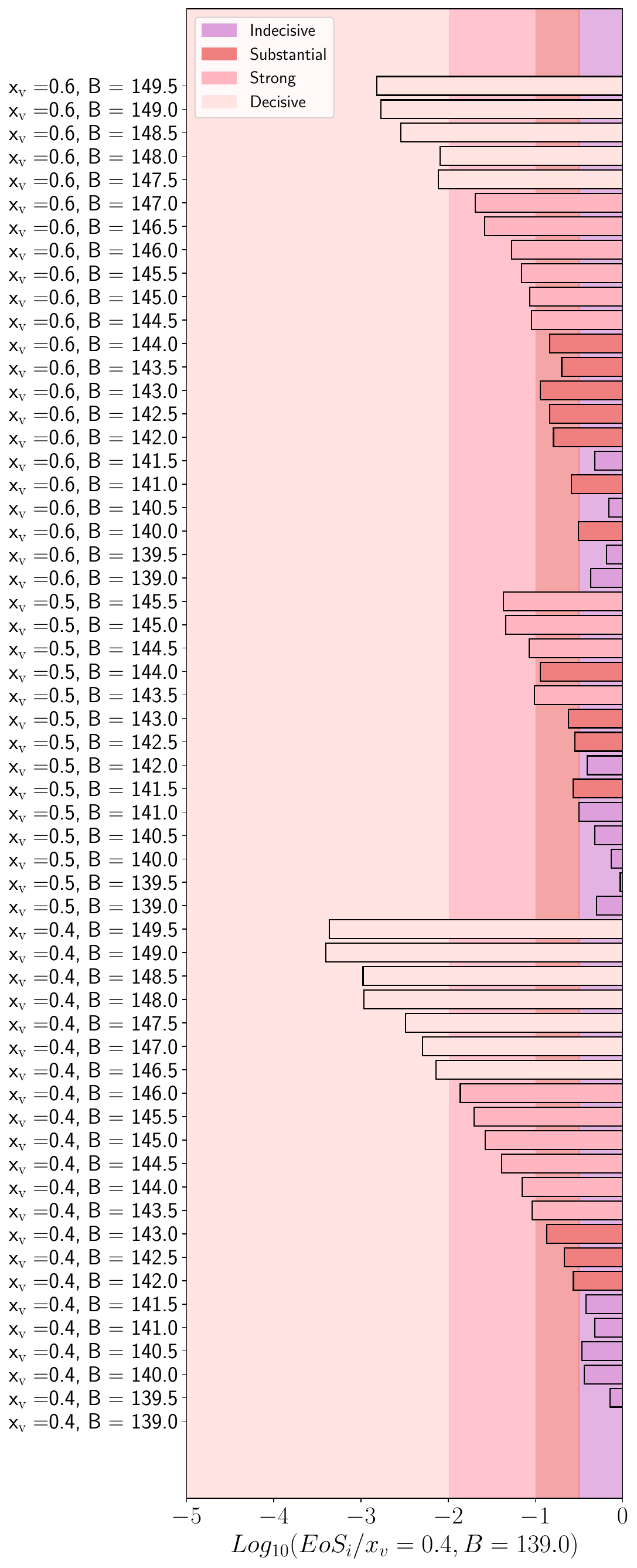}
        \caption{}
        \label{fig:Strange_Odds_onlypulsar}
    \end{subfigure}
     \caption{\raggedright \textbf{(a)}: Odds ratio plot of the hadronic matter EoS {PT(GRDF2-DD2)} with other hadronic matter EoSs. These are the results obtained when HESS J1731-347 was not taken into consideration. The colour scheme is the same as in \cref{fig:Oddsplothadronic}; \textbf{(b)}: Odds ratio plot of the strange matter EoS with $x_v = 0.4$ and B$ = 139.0$ with other strange matter EoSs. These are the results obtained when HESS J1731-347 was not taken into consideration. The colour scheme is the same as in \cref{fig:Oddsplothadronic}. The ticks on the y-axis refer to strange matter EoSs with the corresponding xv and B parameters.}
     \label{fig:Had_SM}
    
\end{figure*}

\begin{figure}[h]
    \flushleft
    \begin{subfigure}{0.55\textwidth}
        \centering
        \includegraphics[width=\textwidth]{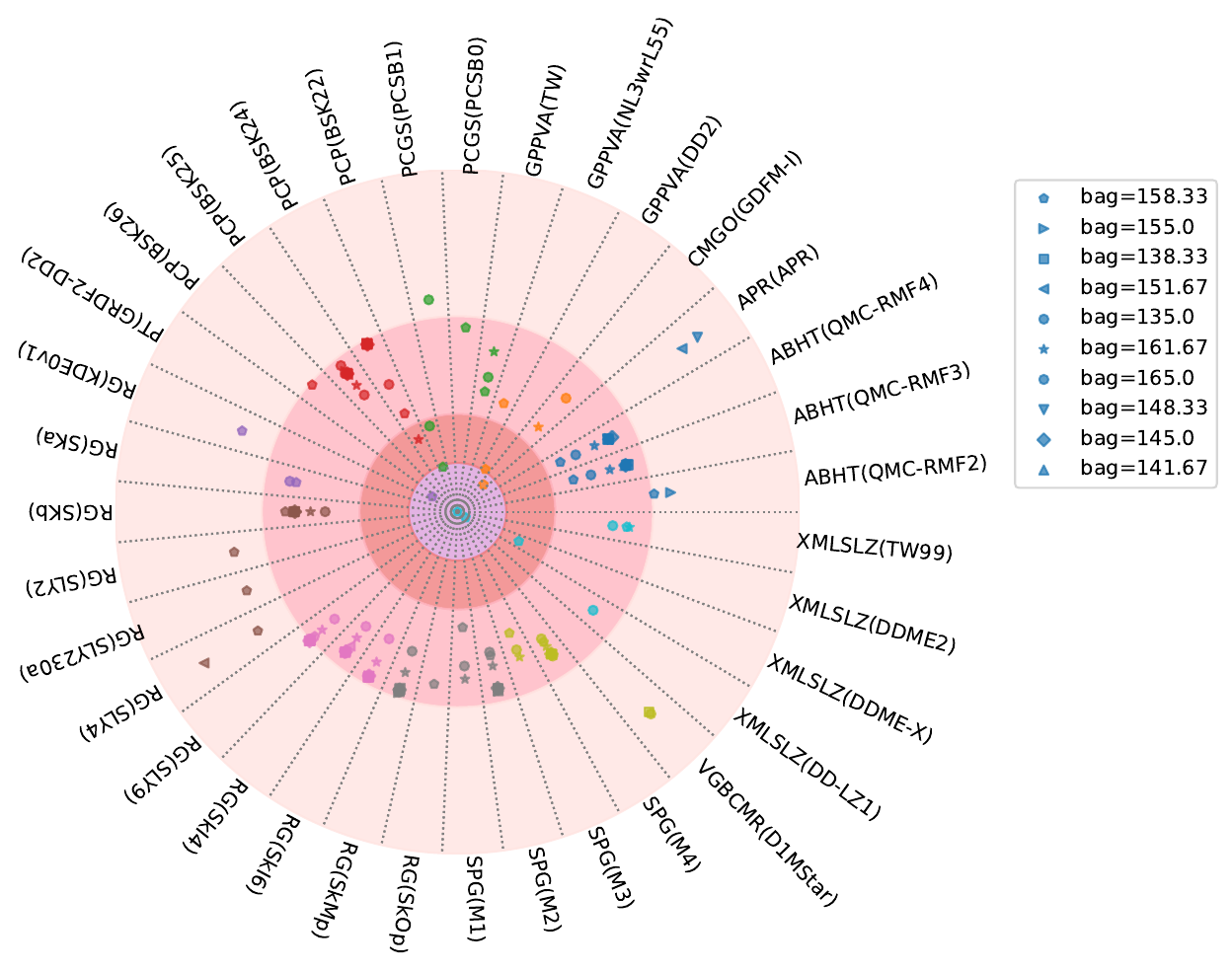}
        \caption{}
        \label{fig:Hybrid_Odds_xv044_onlypulsar}
      
    \end{subfigure}
   \hfill
   \begin{subfigure}{0.55\textwidth}
        \flushleft
        \includegraphics[width=0.82\textwidth]{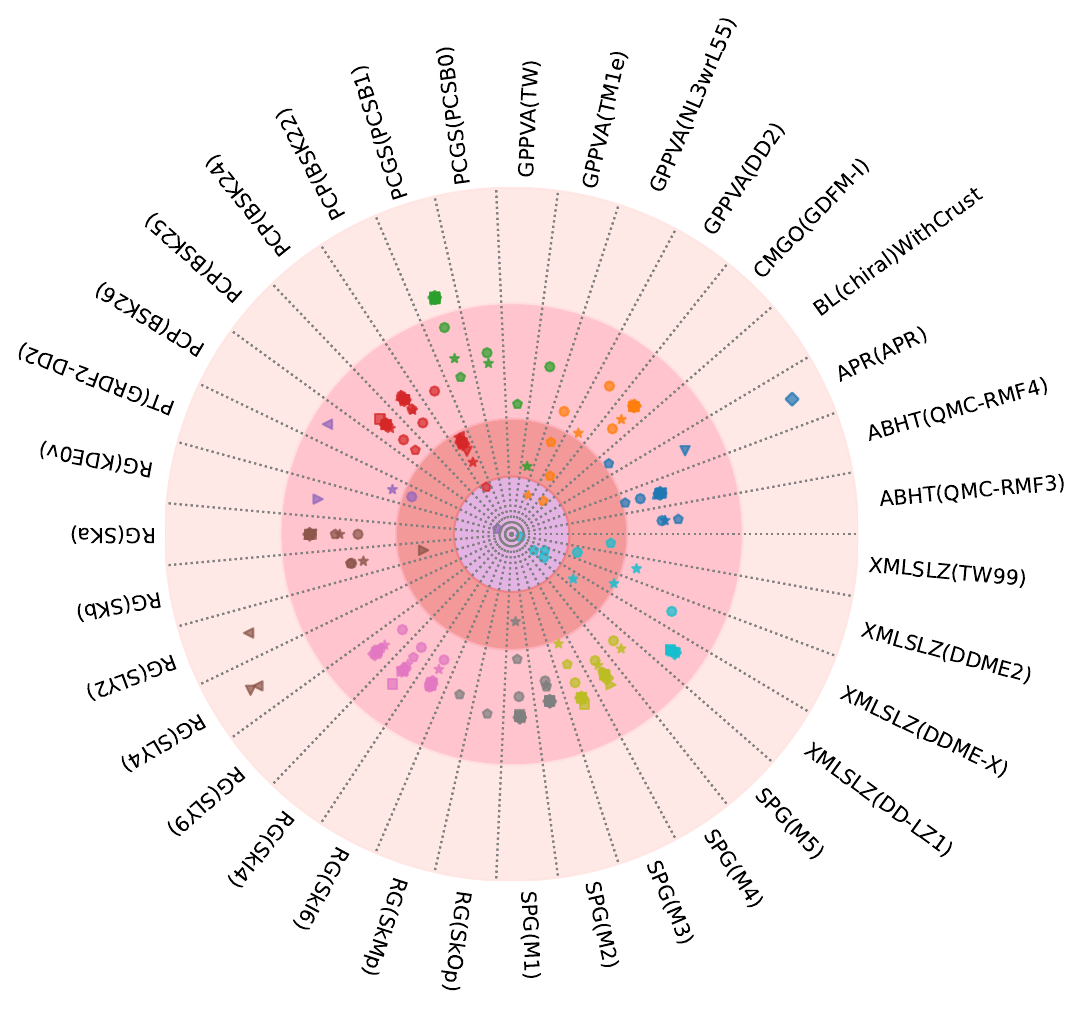}
        \caption{}
        \label{fig:Hybrid_Odds_xv05_onlypulsar}
       
    \end{subfigure}
   \hfill
    \caption{\raggedright The figure illustrates the odds ratio plot obtained when HESS J1731-347 was not considered. \textbf{(a)}: Odds ratio plot of Hyb\_best with EoSs having a fixed $x_v$ value of 0.44; \textbf{(b)}: Odds ratio plot of Hyb\_best with EoSs having a fixed $x_v$ value of 0.5. The nomenclature is the same as \cref{fig:Hybrid_xv044_xv05}.}
    \label{fig:Hybrid_pulsarplot}
\end{figure}

\begin{figure}
    \flushright
    \includegraphics[width=0.9\linewidth]{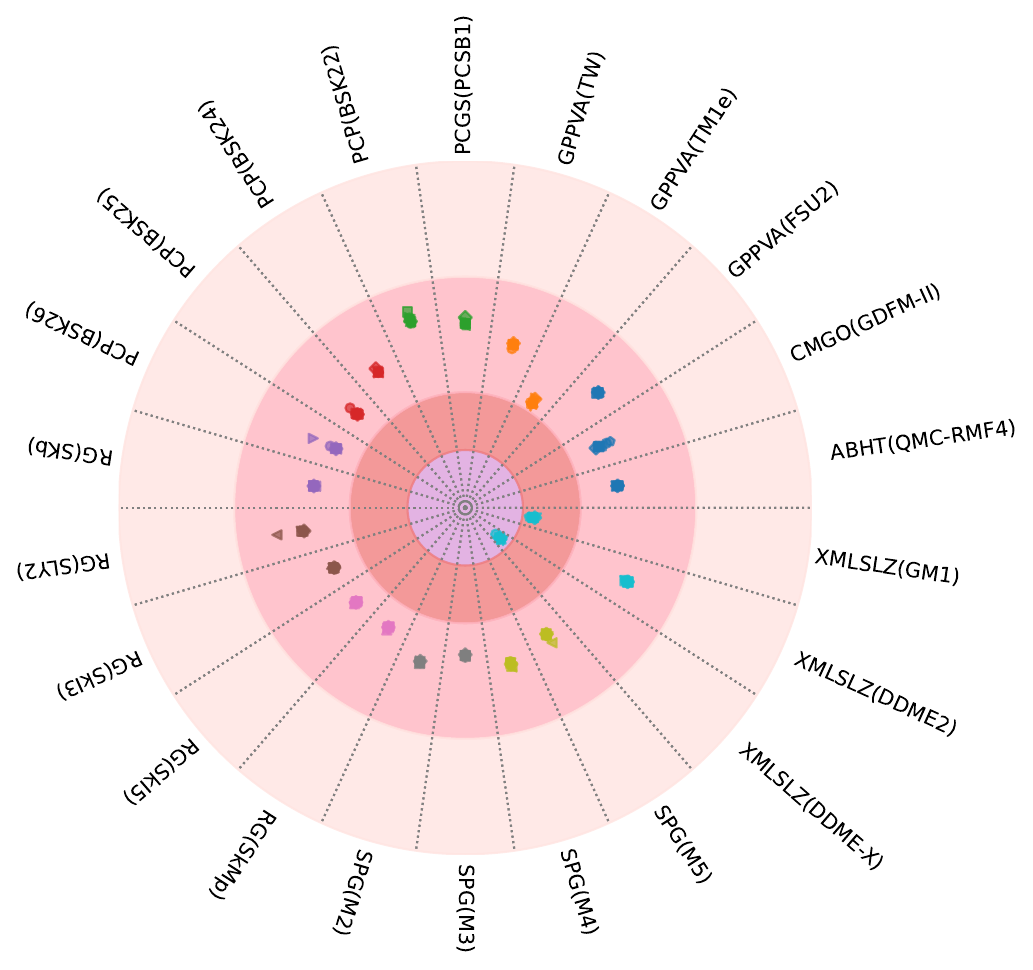}
    \caption{\raggedright The figure illustrates the odds ratio plot of Hyb\_best with EoSs having a fixed $x_v$ value of 0.2 when HESS J1731-347 was not considered. The nomenclature is the same as \cref{fig:Hybrid_xv044_xv05}.}
    \label{fig:Hybrid_xv02_onlypulsar}
\end{figure}

In this section, we discuss the status of the EoSs in the context of astrophysical observations of only PSR J0740+6620, PSR J0030+0451, {PSR J0437-4715}, and GW170817. Only including these observations, we illustrate how each of the hadronic, quark, and hybrid family of EoSs behave in \cref{fig:Had_SM,fig:Hybrid_pulsarplot,fig:Hybrid_xv02_onlypulsar}. In \cref{fig:Had_SM}, we show the odds ratio plot for the hadronic and SM EoSs. From the hadronic family of EoSs, we observe that none of the EoSs could be decisively rejected contrary to the results discussed upon including HESS observation. Furthermore, in this scenario, {PT(GRDF2-DD2)} (following the Brussels-Montreal energy density functional) was the most likely EoS, with {24} other EoSs lying in the 'indecisive' region. This is a significantly greater number than when the data of HESS is considered.\\

The SM EoS analysis shows that although several EoSs can be decisively rejected based only on these observations, the total number of EoSs that can be decisively rejected upon including HESS data significantly increases. In this scenario, the EoS with $x_v$ = 0.4 and B = 139.0 is the most likely, with {15 other} EoSs lying in the 'indecisive' region. Therefore, a similar trend is also followed by comparable equations, where the number of comparable equations decreases when the data of HESS J1731-347 is included. \\

\indent Fig \ref{fig:Hybrid_xv02_onlypulsar} shows the odds ratio plot without considering the data of HESS for a fixed value of $x_v = 0.2$. {On comparing with \cref{fig:Hybrid_xv02}, it} indicates that including the observational data of HESS J1731-347 has significantly helped reject several EoSs (those lying in the `decisive' region). Furthermore, \cref{fig:Hybrid_Odds_xv044_onlypulsar,fig:Hybrid_Odds_xv05_onlypulsar} illustrate the odds ratio plots for the fixed values of $x_v = 0.44\ \rm{and}\ 0.5$ respectively, when HESS is not considered. Contrary to \cref{fig:Hybrid_xv044_xv05}, we see that although several EoSs are still rejected for $x_v = 0.44$, only a handful of EoSs are now rejected for $x_v = 0.5$. One of the most exciting things about this analysis is that Hyb\_best is the most likely EoS independent of whether we include the observation of HESS or not.


\clearpage
\bibliography{main}

\pagebreak
\end{document}